%% file: ch_for_metamat.tex
\documentclass[preprint,12pt]{elsarticle} 

\usepackage{lineno}
\modulolinenumbers[1]
\input{lineno_fix.tex}
\usepackage{color}
\usepackage[usenames,dvipsnames]{xcolor}
\usepackage{fullpage}
\usepackage{amsmath}
\usepackage{amssymb}
\usepackage{mathrsfs}
\usepackage{newfloat}
\usepackage{graphicx}
\usepackage{subfig}
\usepackage{multirow}
\usepackage{stackengine}
\usepackage{caption}
\usepackage{tikz}
\usepackage{multicol}
\usetikzlibrary{arrows,positioning,calc}
\usepackage{pstool}
\usepackage{textcomp}
\usepackage{comment}
\usepackage{array}
\usepackage{pifont}
\usepackage[inline]{enumitem}
\usepackage[hidelinks]{hyperref}
\hypersetup{colorlinks,bookmarksopen,linkcolor=blue,citecolor=blue,urlcolor=blue}

\newcommand{\cmark}{\ding{51}}
\newcommand{\xmark}{\ding{55}}

\definecolor{myblue}{RGB}{0,115,189}
\definecolor{mygreen}{rgb}{0.19,0.61,0.21}

\newcommand{\bs}[1]{{\boldsymbol{#1}}}

\DeclareFloatingEnvironment{algorithm}
\graphicspath{{FigureSven/}}

\journal{Computational Mechanics}

\biboptions{authoryear}

\begin{document}

\begin{frontmatter}
\title{A Comparative Study of Enriched Computational Homogenization Schemes Applied to Two-Dimensional Pattern-Transforming Elastomeric Mechanical Metamaterials\tnoteref{mytitlenote}}
\tnotetext[mytitlenote]{The post-print version of this article is published in \emph{Comput. Mech.}, \href{https://doi.org/10.1007/s00466-023-02428-2}{10.1007/s00466-023-02428-2}. This manuscript version is made available under the \href{https://creativecommons.org/licenses/by/4.0/}{CC-BY 4.0} licence.}

\pagenumbering{arabic}

\author[ME]{S.O.~Sperling}
\ead{S.O.Sperling@tue.nl}

\author[MCS]{T.~Guo}
\ead{T.Guo@tue.nl}

\author[ME]{R.H.J.~Peerlings}
\ead{R.H.J.Peerlings@tue.nl}

\author[ME]{V.G.~Kouznetsova}
\ead{V.G.Kouznetsova@tue.nl}

\author[ME]{M.G.D.~Geers}
\ead{M.G.D.Geers@tue.nl}

\author[ME,CTU]{O.~Roko\v{s}\corref{correspondingauthor}} 
\ead{O.Rokos@tue.nl}

\address[ME]{Mechanics of Materials, Department of Mechanical Engineering, Eindhoven University of Technology, P.O.~Box~513, 5600~MB~Eindhoven, The~Netherlands}
\cortext[correspondingauthor]{Corresponding author.}
\address[MCS]{Centre for Analysis, Scientific Computing and Applications, Department of Mathematics and Computer Science, Eindhoven University of Technology, P.O.~Box~513, 5600~MB~Eindhoven, The~Netherlands}
\address[CTU]{Department of Mechanics, Faculty of Civil Engineering, Czech Technical University in Prague, Th\'{a}kurova~7, 166~29 Prague~6, Czech Republic}

\begin{abstract}
Elastomeric mechanical metamaterials exhibit unconventional behaviour, emerging from their microstructures often deforming in a highly nonlinear and unstable manner. Such microstructural pattern transformations lead to non-local behaviour and induce abrupt changes in the effective properties, beneficial for engineering applications. To avoid expensive simulations fully resolving the underlying microstructure, homogenization methods are employed.
In this contribution, a systematic comparative study is performed, assessing the predictive capability of several computational homogenization schemes in the realm of two-dimensional elastomeric metamaterials with a square stacking of circular holes. In particular, classical first-order and two enriched schemes of second-order and micromorphic computational homogenization type are compared with ensemble-averaged full direct numerical simulations on three examples: uniform compression and bending of an infinite specimen, and compression of a finite specimen.
It is shown that although the second-order scheme provides good qualitative predictions, it fails in accurately capturing bifurcation strains and slightly over-predicts the homogenized response. The micromorphic method provides the most accurate prediction for tested examples, although soft boundary layers induce large errors at small scale ratios. The first-order scheme yields good predictions for high separations of scales, but suffers from convergence issues, especially when localization occurs.

\end{abstract}


\end{frontmatter}


%
%
\section{Introduction}
\label{introduction}
Mechanical metamaterials are characterized by their exotic material properties owing to their engineered structure. Supported by extensive research in recent years, various types of metamaterials hold potential for dedicated applications in distinct fields of engineering, including biomedical, automotive, and aerospace industries, cf.~\citep{Prawoto2012,Krushynska:2023}. An interesting subgroup of such materials, applicable especially in soft robotics and receiving significant scientific attention, is called elastomeric metamaterials. Their extraordinary behaviour relies on the transformation of recurrent patterns within the microstructure during compression. The pattern reconfiguration yields a significant change in effective material properties such as Young's modulus or Poisson's ratio, see, e.g., \cite{Hu2013}, \cite{florijn2014}, \cite{Wu2015}, or~\cite{Coulais2016} and~\cite{Dykstra:202}. One of the earlier works by~\cite{Mullin2007} already showed that the pattern transformation and corresponding change in effective properties is initiated by buckling of the ligaments which separate individual cells.

Pattern transformation of hexagonally shaped cellular structures has been investigated by~\cite{Gao2018}, where two distinct re-configurations were identified. It was shown that the wall thickness to cell size ratio dictates the post-buckling pattern and associated effective characteristics. \cite{Wu2015} evaluated the effect of microstructural buckling on the effective stress--strain path for a triangular structure built from the kagome lattices. Here, the ligaments constructed from mirrored triangles were pre-bent, allowing a smooth structural re-configuration and avoiding abrupt changes in the mechanical response. The in-plane and out-of-plane buckling behaviour of square cellular plates was studied by~\cite{Niknam2018}, where it was concluded that the slenderness ratio, defined as the ratio between the length and the thickness of the specimen, governs the buckling mode and therefore determines pattern switching.

A rigorous numerical study on pattern-transforming mechanical metamaterials is provided by~\cite{Ameen2018b}. The hyper-elastic specimens considered in this study contain a structure of periodically distributed circular holes, which upon compression transform into an arrangement of alternating ellipses. It was found that the post-transformation material behaviour is strongly influenced by the specimen size when the ratio between the specimen height and cell size is relatively small. These size effects are induced by constraints on the pattern transformation close to the sample's fixed edges, which trigger stiff boundary layers. In metamaterial design, as well as predictive modelling involving mechanical metamaterials, one is interested in the effective material properties. Moreover, engineering simulations of components or large structures cannot be done with full account of the microstructure. A homogenized approach is the most efficient solution method for engineering purposes. To retrieve the actual effective material and kinematic behaviour corresponding to a single specimen, \cite{Ameen2018b} employed \textit{ensemble averaging}. To acquire the \textit{ensemble averaged} solution, the material behaviour for a family of translated microstructures is required, meaning that a number of boundary value problems needs to be solved. Since such a brute force approach is computationally expensive, more efficient homogenization techniques are called for. A rich variety of options has been presented in the literature. For a well-characterized microstructure, however, computational homogenization schemes are typically assumed to be among the most accurate approaches~\citep{Geers2010}. In spite of its high accuracy, standard first-order computational homogenization cannot capture size effects, since the scale separation assumption does not allow for non-local behaviour~\citep{Ameen2018a}. Enriched computational homogenization schemes are thus required to achieve a higher accuracy, especially for specimens with small differences between the length of the microstructural features and considered macrostructural specimen size.

In order to reflect for non-locality and size effects, second-order computational homogenization~\citep{Kouznetsova2002,Kouznetsova2004} incorporates a generalized continuum formulation at the macro-scale, which makes it a serious candidate for elastomeric mechanical metamaterials. \cite{Anthoine2010} applied second-order homogenization to functionally graded materials to quantify the effect of microstructural grading. The intrinsic length scale which influences the higher-order stiffness can be directly related to the graded microstructure of the material. \cite{Forest2010} used the same method to investigate the size effect of Representative Volume Elements (RVEs) for elastic and elastoplastic composites subjected to quadratic Dirichlet boundary conditions. The research conducted by~\cite{Nguyen2014} studied the behaviour of elastic and elastoplastic materials containing a cellular hexagonal microstructure through second-order computational homogenization. Via comparison to full numerical simulations, it was concluded that this approach accurately predicts the behaviour of structures consisting of repeated hexagonal patterns, but the method fails for high degrees of macroscopic imperfections. Generalization of second-order homogenization method to include body forces avoiding spurious effects were carried out in~\citep{Luscher:2010,Blanco:2016,Yvonnet:2020}, and in the context of non-linear cellular metamaterials in~\citep{Wu2023}. An overview comparison of second-order models has been presented by~\cite{Lopes:2022,Lopes:2022a}.

Another suitable option for homogenization of elastomeric metamaterials is micromorphic homogenization scheme introduced by~\cite{Rokos2018}, which incorporates kinematic coupling between individual cells through an additional field variable. The essence of the adopted approach lies in the decomposition of the displacement field into slow and fast varying components. In addition, a predefined fluctuation field corresponding to the pattern transformation is introduced, representing prior knowledge about the most important microstructural deformation modes. The method has been compared to ensemble averaged results for uniformly compressed structures by~\cite{Ameen2018b}, where the accuracy of the method has been assessed for a macroscopically one-dimensional problem. The methodology has been generalized to multiple micromorphic fields in~\citep{Rokos2019a}, and multiscale buckling addressed in~\citep{Bree2019}.

The aim of this paper is to compare the performance of the two above-mentioned enriched computational homogenization schemes (i.e., the second-order and micromorphic) in the context of elastomeric mechanical metamaterials exhibiting pattern transformations against the reference ensemble-averaged solution on three examples considered for simplicity in two-dimensional setting under plane strain assumption. Both methods include an enhanced continuum description at the macro-scale, allowing to account for relevant size and boundary effects. For completeness, results corresponding to the scale-independent first-order computational homogenization scheme are included as well.

The remainder of this paper is organized as follows. First, Section~\ref{sec:Material} specifies the elastomeric metamaterial considered, through its geometry, constitutive laws, and basic mechanical behaviour. Section~\ref{sec:Methodology} introduces the reference solution represented by the ensemble average of Direct Numerical Simulations~(DNS) corresponding to a family of translated microstructures. The classical first-order computational homogenization method is then briefly reviewed, followed by a description of the second-order and micromorphic homogenization schemes. In Section~\ref{sec:Results}, the considered load cases are presented, and the results obtained with the different homogenization methods are discussed. Finally, overall summary and conclusions are provided in Section~\ref{sec:Conclusion}.

Throughout this paper, the following notational conventions are adopted: scalars~$a$, vectors~$\vec{a} = a_{i}\vec{e}_{i}$, second-order tensors~$\boldsymbol{A} = A_{ij}\vec{e}_{i}\vec{e}_{j}$, third-order tensors~${}^{3}\boldsymbol{A} = A_{ijk}\vec{e}_{i}\vec{e}_{j}\vec{e}_{k} $, scalar product~$\vec{a} \cdot \vec{b} = a_{i}b_{i}$, single contraction~$\boldsymbol{A} \cdot \vec{b} = A_{ij}b_{j}\vec{e}_{i}$, double contraction~$\boldsymbol{A} : \boldsymbol{B} = A_{ij}B_{ji}$, transpose~$\boldsymbol{A}^{\mathrm{T}}, A_{ij}^{\mathrm{T}} = A_{ji}$, gradient operator~$\vec{\nabla} \vec{a} = \frac{\partial a_{j}}{\partial X_{i}} \vec{e}_{i} \vec{e}_{j}$, and divergence operator~$\vec{\nabla} \cdot \vec{a} = \frac{\partial a_{i}}{\partial X_{i}}$. Unless indicated otherwise, summation over repeated indices is assumed for tensor operations.
%
%
\section{Metamaterial Specification}
\label{sec:Material}

\begin{figure}
	\centering
	\subfloat[cellular structure]{\includegraphics[height=5cm]{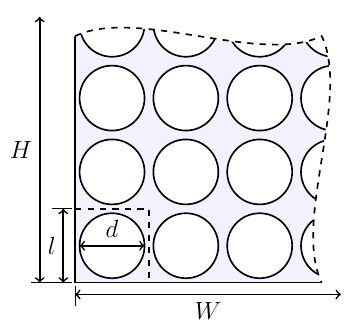}\label{fig:Schematic_a}}
	\subfloat[stress--strain diagram]{\includegraphics[height = 5cm]{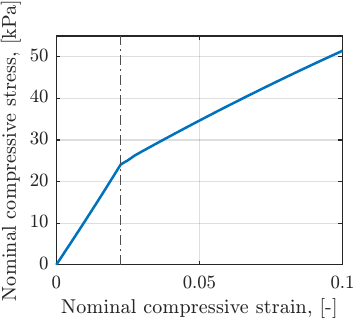}\label{fig:Schematic_b}}
    \subfloat[deformation pattern]{\raisebox{0.3cm}{\includegraphics[trim = {1cm 1cm 0.2cm 1cm},clip,height=4.5cm]{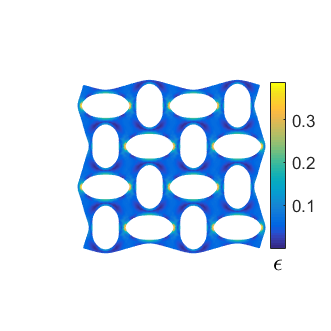}\label{fig:PatTrans_b}}}
	\caption{(a) Sketch of the two-dimensional elastomeric metamaterial, including the geometrical parameters, and~(b) corresponding stress--strain response in compression. (c)~Deformation pattern of a~$4 \times 4$ periodic specimen subjected to~$7.5\%$ of overall compressive strain; the colour indicates the norm of the Green--Lagrange strain~$\boldsymbol{E}$, defined as~$\epsilon = \sqrt{\boldsymbol{E}:\boldsymbol{E}}$.}
	\label{fig:Schematic}
\end{figure}

The geometry considered in this study is inspired by \textit{Specimen 1}, as described in~\citep{Bertoldi2008}, which was adopted also in the study of~\cite{Ameen2018b}, consisting of a square stacking of circular holes within a two-dimensional domain subject to plane strain conditions. A schematic representation of the cellular structure is provided in Fig.~\ref{fig:Schematic_a}, illustrating the underlying square unit cell, its size~$l$, and hole diameter~$d$. In this study, the adopted geometrical values are set as~$l = 9.97$ mm and~$d = 8.67$ mm. The specimen's width and height are denoted by~$W$ and~$H$ and depend on the amount of cells in the horizontal or vertical direction. Upon application of a compressive load, typically a bi-linear stress--strain diagram is obtained, as shown in Fig.~\ref{fig:Schematic_b}. The diagram splits into two parts by the vertical dashed line, indicating the bifurcation strain of the microstructure. At this strain, individual circular voids transform abruptly into ellipses with alternating horizontal and vertical major axes, resulting in a significant drop of the effective stiffness. The pattern transformation and resulting mechanical response have been thoroughly investigated, e.g., by~\cite{Mullin2007}, \cite{Bertoldi2008}, \cite{Hu2013}, or~\cite{Coulais2016}. Fig.~\ref{fig:PatTrans_b} illustrates a~$4 \times 4$ part of an infinite specimen, subjected to~$7.5\%$ overall vertical compressive strain. At that strain, the specimen is in the post-bifurcation regime, i.e., the pattern is in the transformed state and the material's mechanical response is now related to the new microstructural setting (cf. Fig.~\ref{fig:Schematic_b}). When replacing the periodicity conditions imposed on the~$4 \times 4$ subsection by prescribed displacements, the fluctuating edges resulting from the pattern transformation will be constrained. This explains the mechanism behind the formation of stiff boundary layers, dictating the mechanical response for specimens containing a low ratio between the specimen height and cell length ($H/l$). This ratio will be referred to as the scale ratio in what follows.

The elastomeric base material, employed throughout all examples discussed in Section~\ref{sec:Results}, is a hyperelastic material. The strain energy density is described by a two-term~$I_{1}$-based compressible Mooney--Rivlin model, which in a two-dimensional setting reads as
\begin{equation}
W(\boldsymbol{F}(\vec{X})) = c_{1}(I_{1}-2) + c_{2}(I_{1} - 2)^{2} - 2c_{1}\mathrm{log}J + \frac{K}{2}(J - 1)^{2},
\label{eq:StrainDensity}
\end{equation}
where~$I_{1}$ indicates the first invariant of the right Cauchy--Green tensor~$\boldsymbol{C} = \boldsymbol{F}^{\mathrm{T}}\cdot \boldsymbol{F}$, and~$J = \det{(\bs{F})}$ represents the volume change ratio. Constants~$c_{1}$, $c_{2}$, and~$K$ are constitutive parameters, set to~$c_{1} = 0.55$~MPa, $c_{2} = 0.3$~MPa, and~$K= 55$~MPa, according to the experimental data of~\cite{Bertoldi2008}.
%
%
\section{Homogenization Methodologies}
\label{sec:Methodology}
%
%
\subsection{Reference Solution Based on Ensemble Averaging}
\label{sec:ensemble}
In the case of heterogeneous materials, the DNS results typically reveal highly oscillatory kinematic fields, which tend to shift periodically with the microstructural position relative to the macroscopic sample. To obtain a representative reference solution to assess the accuracy of computational homogenization methods, an ensemble averaged DNS solution of a family of translated microstructures is established. Averaging the deformation fields over all translated microstructures filters out the fast oscillations and results in a macroscopically smooth solution. For constructing this reference solution, each translated microstructure is assumed to have equal probability of occurrence, as employed by, e.g., \cite{Drugan}, \cite{Smyshlyaev}, or~\cite{Ameen2018a}, relying on the assumption that the exact positioning of the microstructure is not known a-priori.

\begin{figure}
	\centering
	\subfloat[microstructural translation]{\raisebox{1.1cm}{\includegraphics{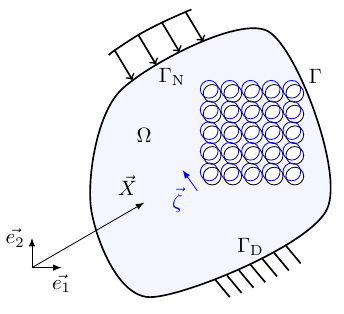}\label{fig:GeneralMethods_a}}}
	\subfloat[first-order computational homogenization]{\includegraphics{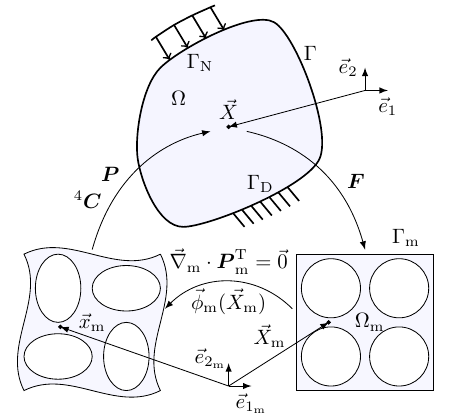}\label{fig:GeneralMethods_b}}
	\caption{Sketch of a macroscopic problem domain~$\Omega$ with boundary~$\Gamma$; Dirichlet and Neumann boundary conditions are applied on boundary segments~$\Gamma_{\mathrm{D}}$ and~$\Gamma_{\mathrm{N}}$. (a)~Illustration of a microstructural translation~$\vec{\zeta}$ (in blue) with respect to the reference configuration (black). The DNS domain~$\Omega^\mathrm{dns}(\vec{\zeta})\subseteq\Omega$ contains all (shifted) holes, whereas~$\Omega$, which is used in homogenization approaches, is in this case simply connected and ignores the underlying microstructure. (b)~A sketch of the computational homogenization scheme; the microscale problem in the reference (bottom right) and deformed (bottom left) configuration considered at a macroscopic point of interest~$\vec{X}$.}
	\label{fig:GeneralMethods}
\end{figure}

In Fig.~\ref{fig:GeneralMethods_a}, a schematic representation of the macroscopic specimen in the reference configuration is given, with domain~$\Omega$, outer boundary~$\Gamma$, and position vector~$\vec{X}$. The vector~$\vec{\zeta}$ is related to the translated positions of the microstructure with respect to its reference position. For the ensemble averaged (reference) solution, one needs to average over the recurrent periodic pattern, which consists of~$2\times2$ cells in the considered case. Consequently, the domain including the required microstructural translations spans~$\vec{\zeta} \in Q = [-l,l] \times [-l,l]$. The reference solution is then obtained as
\begin{equation}
\vec{\overline{u}}(\vec{X}) = \frac{1}{|Q|}\int_{Q}\vec{u}(\vec{X},\vec{\zeta})\,\mathrm{d}\vec{\zeta},
\label{eq:EnsembleAverage}
\end{equation}
where~$\vec{u}(\vec{X},\vec{\zeta})$ denotes the displacement field corresponding to the microstructure translated by~$\vec{\zeta}$, governed by the classical balance of linear momentum, i.e.,
\begin{equation}
\vec{\nabla}\cdot\bs{P}^\mathrm{T}(\vec{X},\vec{\zeta}) = \vec{0}, \quad \vec{X} \in \Omega^\mathrm{dns}(\vec{\zeta}) \subseteq \Omega,
\label{eq:dns}
\end{equation}
valid over the entire domain with fully resolved microstructure containing all holes shifted by~$\vec{\zeta}$, $\vec{X} \in \Omega^\mathrm{dns}(\vec{\zeta}) \subseteq \Omega$, cf. Fig.~\ref{fig:GeneralMethods_a}. Because the static behaviour is of prime interest herein, all inertia effects (as well as body forces) are neglected throughout. The reference homogenized solution~$\vec{\overline{u}}(\vec{X})$ is for convenience computed numerically by discretizing the integral of Eq.~\eqref{eq:EnsembleAverage} into a set of grid points spanning~$Q$, and calculating the DNS solution~$\vec{u}(\vec{X},\vec{\zeta})$ for each realization.
%
%
\subsection{First-Order Computational Homogenization}
Computational homogenization is a multiscale technique, which incorporates detailed microstructural information by calculating the microscale and macro-scale boundary value problem in a nested manner~\citep{Geers2010}. This method provides a powerful numerical tool for macroscopically homogeneous and microscopically heterogeneous materials, for which the macroscopic constitutive law is not available in a closed form, but implicitly through the solution of an underlying microstructural problem. The macroscopic first Piola--Kirchhoff stress tensor can be expressed as
\begin{equation}
\boldsymbol{P} = \boldsymbol{f}\big(\vec{X},\boldsymbol{P}_{\mathrm{m}}\big(\boldsymbol{F}(\vec{X})\big)\big), \quad \vec{X} \in \Omega,
\label{eq:constLaw}
\end{equation}
where~$\boldsymbol{F}$ denotes the macroscopic deformation gradient and~$\boldsymbol{P}_{\mathrm{m}}$ the microscopic first Piola--Kirchhoff stress tensor. From Eq.~\eqref{eq:constLaw} it is clear that the macroscopic stress is related to the microscopic stress resulting from the macroscopic deformation. In Fig.~\ref{fig:GeneralMethods_b} a schematic representation of the computational homogenization concept applied to elastomeric mechanical metamaterials is given. Note that two nested boundary value problems are involved: (i)~a macroscopic problem (Fig.~\ref{fig:GeneralMethods_b} top) governed by the classical balance equation, i.e.,
\begin{equation}
\vec{\nabla} \cdot \boldsymbol{P}^{\mathrm{T}}(\vec{X}) = \vec{0}, \quad \vec{X} \in \Omega,
\label{eq:M-balance}
\end{equation}
and~(ii) a microscopic problem (Fig.~\ref{fig:GeneralMethods_b} bottom) with a similar balance equation
\begin{equation}
\vec{\nabla}_{\mathrm{m}} \cdot \boldsymbol{P}_{\mathrm{m}}^{\mathrm{T}}(\vec{X}_{\mathrm{m}}) = \vec{0}, \quad \vec{X}_{\mathrm{m}} \in \Omega_{\mathrm{m}},
\label{eq:balancemicro}
\end{equation}
where inertia and body forces have been dropped again in both cases, cf. Eq.~\eqref{eq:dns}. Note that unlike the DNS problem of Eq.~\eqref{eq:dns}, the macroscopic problem of Eq.~\eqref{eq:M-balance} is solved over the (simply connected) domain~$\Omega$, neglecting the underlying metamaterial microstructure. In Eq.~\eqref{eq:balancemicro}, $\vec{X}_{\mathrm{m}}$ denotes the microscopic spatial position vector and~$\Omega_{\mathrm{m}}$ the microscopic domain. Recall that, throughout this paper, the macroscopic~$\vec{\nabla}$ and microscopic~$\vec{\nabla}_{\mathrm{m}}$ differential operators are defined with respect to the undeformed configurations.

To obtain the macroscopic first Piola--Kirchhoff stress tensor~$\boldsymbol{P}$ at a point of interest~$\vec{X}$, the corresponding macroscopic deformation gradient tensor~$\boldsymbol{F}({\vec{X}})$ is imposed on the microscopic problem. The relation between the microscopic deformation field and macroscopic deformation gradient is given by
\begin{equation}
\vec{x}_{\mathrm{m}} = \boldsymbol{F}\cdot \vec{X}_{\mathrm{m}} + \vec{w}(\vec{X}_\mathrm{m}),
\label{eq:TaylorExpansionFirst}
\end{equation}
which corresponds to a Taylor expansion of the macroscopic deformation field truncated after its leading term. In Eq.~\eqref{eq:TaylorExpansionFirst}, $\vec{x}_{\mathrm{m}}$ and~$\vec{X}_{\mathrm{m}}$ are the position vectors in the deformed and reference configuration of the microscopic problem, whereas~$\vec{w}$ represents the local zero-mean micro-fluctuation correction field. The boundary conditions imposed on the microscopic problem, also referred to as Representative Volume Element~(RVE), are a combination of periodicity conditions,
\begin{equation}
\vec{w}(\Gamma_{\mathrm{m}_{\mathrm{T}}}) = \vec{w}(\Gamma_{\mathrm{m}_{\mathrm{B}}})\quad \text{and}\quad \vec{w}(\Gamma_{\mathrm{m}_{\mathrm{L}}}) = \vec{w}(\Gamma_{\mathrm{m}_{\mathrm{R}}}),
\label{eq:pbcs}
\end{equation}
and prescribed displacements resulting from the macroscopic deformation gradient,
\begin{equation}
\vec{x}_{\mathrm{m},i} = \boldsymbol{F} \cdot \vec{X}_{\mathrm{m},i} \quad \text{for} \quad i = 1,2,4,
\label{eq:dirbcs}
\end{equation}
see Fig.~\ref{fig:SecondOrderMicroDomain}. In Eq.~\eqref{eq:pbcs}, $\Gamma_{\mathrm{m}_\bullet}$ denotes individual RVE boundary segments, and~$\bullet$ represents right~(R), left~(L), top~(T), and bottom~(B) RVE boundaries. Spatial positions~$\vec{X}_{\mathrm{m},i}$, $i = 1,2,4$, denote in Eq.~\eqref{eq:dirbcs} three RVE corner points, as indicated in Fig.~\ref{fig:SecondOrderMicroDomain} where a schematic representation of a rectangular RVE domain is shown. Note that equivalently to fixing the three control points in Eq.~\eqref{eq:dirbcs}, $\vec{w}$ may be considered periodic on top of~$\boldsymbol{F}\cdot\vec{X}_\mathrm{m}$ while being orthogonal with respect to a constant inside the RVE domain~$\Omega_\mathrm{m}$ to remove rigid body motions~\cite[see, e.g.,][]{Miehe:2007}.

After solving the microscopic RVE problem according to Eq.~\eqref{eq:balancemicro}, the macroscopic stress at a material point~$\vec{X}$ is obtained by volume averaging the microscopic stress field over~$\Omega_{\mathrm{m}}$, unlike ensemble averaging considered for the underlying solution of Section~\ref{sec:ensemble}, i.e.,
\begin{equation}
\boldsymbol{P}(\vec{X}) = \frac{1}{|\Omega_{\mathrm{m}}|} \int_{\Omega_{\mathrm{m}}} \boldsymbol{P}_{\mathrm{m}}\,\mathrm{d}\Omega_{\mathrm{m}}, \quad \vec{X} \in \Omega,
\label{eq:macroscopicstressP}
\end{equation}
which is consistent with the Hill--Mandel condition. This process can be repeated for each macroscopic point of interest. Generally, the micro- and macro-boundary value problems are conveniently solved using the Finite Element Method, where each macroscopic integration point in~$\Omega$ is related to a RVE. The macroscopic problem of Eq.~\eqref{eq:M-balance} can be solved using a standard Newton iterative algorithm, resulting in an integrated scheme in which the micro- and macro-boundary value problems are solved in a nested manner.
%
%
\subsection{Second-Order Computational Homogenization}
The second-order homogenization method may be regarded as an extension of its first-order counterpart, relying on a Taylor series expansion of the deformation field truncated after the second term. The local deformation field within a RVE then takes the form (cf. Eq.~\eqref{eq:TaylorExpansionFirst})
\begin{equation}
\vec{x}_{\mathrm{m}} = \boldsymbol{F}\cdot \vec{X}_{\mathrm{m}} + \frac{1}{2}\vec{X}_{\mathrm{m}}\cdot{}^{3}\boldsymbol{G}\cdot \vec{X}_{\mathrm{m}} + \vec{w}(\vec{X}_\mathrm{m}),
\label{eq:TaylorExpansion}
\end{equation}
where the second term on the right-hand side accounts for non-local strain gradient effects through the gradient of the deformation gradient, $^{3}\bs{G} = \vec{\nabla} \bs{F}$. Such a formulation requires an extended continuum framework, which is here given by Mindlin's strain gradient continuum at the macro level, cf., e.g., \cite{Mindlin1965} or~\cite{Mindlin1968}. Following the lines of reasoning presented by~\cite{Kouznetsova2004}, a classical continuum of Eq.~\eqref{eq:balancemicro} can be adopted at the microscale level.
\begin{figure}
	\centering
	\includegraphics{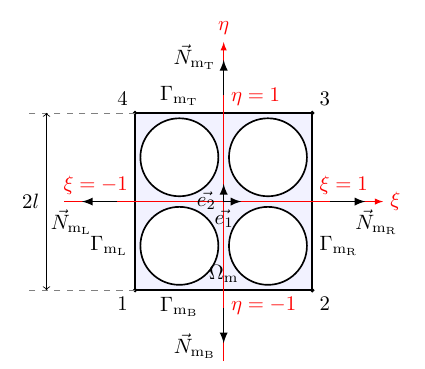}
	\caption{Schematic representation of the adopted square RVE with domain~$\Omega_{\mathrm{m}}$, consisting of~$ 2 \times 2 $ unit cells. The RVE has four corner points, $1, \dots, 4$, four boundary segments~$\Gamma_{\mathrm{m}_\bullet}$ with associated outer normals~$\vec{N}_{\mathrm{m}_\bullet}$, where~$\bullet$ represents right~(R), left~(L), top~(T), and bottom~(B) edges, and is parametrized by two normalized coordinates~$\xi \in [-1,1]$ and~$\eta \in [-1,1]$.}
	\label{fig:SecondOrderMicroDomain}
\end{figure}

Introducing the higher-order stress, $^3\boldsymbol{Q}$, which is work conjugate to~$^{3}\boldsymbol{G}$, the non-classical balance equation reads
\begin{equation}
\vec{\nabla} \cdot \big(\boldsymbol{P}^{\mathrm{T}} - (\vec{\nabla} \cdot {}^3\boldsymbol{Q})^{\mathrm{T}}\big) = \vec{0}, \quad \vec{X} \in \Omega.
\label{eq:2nd:balance}
\end{equation}
In Eq.~\eqref{eq:2nd:balance}, consistently with the previous considerations any body and inertia forces have been neglected, as well as Neumann boundary conditions for simplicity. The presence of higher-order derivatives in Eq.~\eqref{eq:2nd:balance} requires~$C^{1}$ continuity of the displacement field over the entire domain~$\Omega$ to obtain the gradient of the deformation gradient at every point~\citep{Lesicar:2014}. From a finite element perspective such a requirement poses an additional complication, since the conventional shape functions typically possess~$C^{0}$ continuity only. In what follows, this is addressed by resorting to a mixed finite element formulation, which introduces a so-called relaxed deformation gradient field~$\widehat{\bs{F}}$. The deformation gradient, containing gradient of the displacement field, i.e., $ \bs{F} = \bs{I} + (\vec{\nabla}\vec{u})^\mathrm{T}$, and the relaxed deformation gradient field denoted~$\widehat{\bs{F}}$ are coupled via Lagrange multipliers, requiring that~$ \widehat{\bs{F}} = \bs{F} $ holds in the weak sense; for more details see~\citep{Xia1996,Kouznetsova2004,Luscher:2010}. Within such a mixed formulation, choice of combinations of integration rules and shape functions for the displacement, Lagrange multiplier, and relaxed deformation gradient fields are of high importance. This is due to potential instabilities polluting the solution, which may occur and significantly deteriorate accuracy of obtained results. To this end, in the numerical examples employed below in Section~\ref{sec:Results}, stable combinations have been employed~\cite[see, e.g.,][Section~4.3]{Kouznetsova2004}, with no spurious instabilities observed.

In addition to the macroscopic first Piola--Kirchhoff stress tensor, specified in Eqs.~\eqref{eq:constLaw} and~\eqref{eq:macroscopicstressP}, the higher-order macroscopic stress tensor is computed from each RVE solution as
\begin{equation}
{}^{3}\boldsymbol{Q} = \frac{1}{2| \Omega_{\mathrm{m}}|}\int_{\Omega_{\mathrm{m}}}\big(\boldsymbol{P}^{\mathrm{T}}_{\mathrm{m}}\vec{X}_{\mathrm{m}} + \vec{X}_{\mathrm{m}}\boldsymbol{P}_{\mathrm{m}}\big)\,\mathrm{d}\Omega_{\mathrm{m}},
\end{equation}
which does not involve higher-order microscopic stresses since the RVE obeys a classical continuum description. This choice effectively circumvents the need for a higher-order characterization of the microstructural problem.

However, to enforce the gradient effects occurring at the macroscopic scale on the micro problem, specific RVE boundary conditions are required, as elaborated in~\citep{Kouznetsova2004}. The starting point for deriving the microscopic RVE boundary conditions is Eq.~\eqref{eq:TaylorExpansion}, which provides the microscopic deformation gradient in the form
\begin{equation}
\boldsymbol{F}_{\mathrm{m}} = (\vec{\nabla}_{\mathrm{m}}\vec{x}_{\mathrm{m}})^{\mathrm{T}} = \boldsymbol{F} + \vec{X}_{\mathrm{m}} \cdot {}^3\boldsymbol{G} + (\vec{\nabla}_{\mathrm{m}}\vec{w})^{\mathrm{T}}.
\end{equation}
To conform to the classical averaging theorem given by
\begin{equation}
\boldsymbol{F} = \frac{1}{|\Omega_\mathrm{m}|}\int_{\Omega_{\mathrm{m}}}\boldsymbol{F}_{\mathrm{m}}\,\mathrm{d}\Omega_{\mathrm{m}},
\end{equation}
two requirements have to be satisfied. First,
\begin{equation}
\frac{1}{|\Omega_{\mathrm{m}}|}\int_{\Omega_{\mathrm{m}}}\vec{X}_{\mathrm{m}}\,\mathrm{d}\Omega_{\mathrm{m}} = \vec{0},
\end{equation}
which is guaranteed by simply shifting the coordinate system to the RVE centre. The second requirement reads
\begin{equation}
\frac{1}{|\Omega_{\mathrm{m}}|}\int_{\Omega_{\mathrm{m}}}(\vec{\nabla}\vec{w})^{\mathrm{T}}\,\mathrm{d}\Omega_{\mathrm{m}} = \frac{1}{|\Omega_{\mathrm{m}}|}\int_{\Gamma_{\mathrm{m}}}\vec{w}\vec{N}_{\mathrm{m}}\,\mathrm{d}\Gamma_{\mathrm{m}} = \boldsymbol{0},
\label{eq:Condition2}
\end{equation}
where the divergence theorem has been used to transform the volume integral into a contour integral along~$\Gamma_{\mathrm{m}}$, with~$\vec{N}_{\mathrm{m}}$ denoting its outward unit normal, cf. Fig.~\ref{fig:SecondOrderMicroDomain}. Assuming~${}^3\bs{G} = {}^3\bs{0}$ yields~$\vec{N}_{\mathrm{m}_{\mathrm{T}}}(\Gamma_{\mathrm{m}_{\mathrm{T}}}) = -\vec{N}_{\mathrm{m}_{\mathrm{B}}}(\Gamma_{\mathrm{m}_{\mathrm{B}}})$ and~$\vec{N}_{\mathrm{m}_{\mathrm{R}}}(\Gamma_{\mathrm{m}_{\mathrm{R}}}) = -\vec{N}_{\mathrm{m}_{\mathrm{L}}}(\Gamma_{\mathrm{m}_{\mathrm{L}}})$, which in turn reduces the requirement of Eq.~\eqref{eq:Condition2} to the standard periodicity condition of Eq.~\eqref{eq:pbcs}. When~${}^3\boldsymbol{G}\neq {}^3\boldsymbol{0}$, however, a non-periodic deformation results, requiring periodicity in a generalized sense, defined as
\begin{align}
\vec{x}_{\mathrm{m}_{\mathrm{R}}} & = \vec{x}_{\mathrm{m}_{\mathrm{L}}} + 2l\boldsymbol{F}\cdot \vec{N}_{\mathrm{m}_{\mathrm{R}}} + 2l^2\eta\vec{N}_{\mathrm{m}_{\mathrm{R}}} \cdot {}^3\boldsymbol{G} \cdot \vec{N}_{\mathrm{m}_{\mathrm{T}}}, \label{eq:genperiodic1} \\
\vec{x}_{\mathrm{m}_{\mathrm{T}}} & = \vec{x}_{\mathrm{m}_{\mathrm{B}}} + 2l\boldsymbol{F} \cdot \vec{N}_{\mathrm{m}_{\mathrm{T}}} + 2l^2\xi\vec{N}_{\mathrm{m}_{\mathrm{T}}} \cdot {}^3\boldsymbol{G} \cdot \vec{N}_{\mathrm{m}_{\mathrm{R}}}, \label{eq:genperiodic2}
\end{align}
where~$\xi$ and~$\eta$ are normalized RVE coordinates, cf. Fig.~\ref{fig:SecondOrderMicroDomain}. The four corner points are displaced according to Eq.~\eqref{eq:TaylorExpansion} in which~$\vec{w}(\vec{X}_{\mathrm{m},i}) = \vec{0}$, $i = 1,\dots,4$, similarly to Eq.~\eqref{eq:dirbcs}. Since this requirement may lead to stress concentrations close to RVE corner points~\citep{Lesicar:2014,Lopes:2022}, prescribed corner point displacements may be relaxed and replaced by a condition requiring orthogonality of~$\vec{w}$ with respect to a constant~\citep{Luscher:2010}. In the examples that follow, the former option according to~\cite{Kouznetsova2002} has been adopted. Finally, to relate the microscopic variables to the macroscopic gradient of the deformation gradient, the following conditions must hold
\begin{equation}
\int_{\Gamma_{\mathrm{m_{L}}}}\vec{w}\,\mathrm{d}\Gamma_{\mathrm{m}} = \vec{0} \quad \text{and} \quad \int_{\Gamma_{\mathrm{m_{B}}}}\vec{w}\,\mathrm{d}\Gamma_{\mathrm{m}} = \vec{0},
\label{eq:fluctmeanzero}
\end{equation}
which are consistent with the generalized periodicity of~$\vec{w}$ on the right and top boundaries, i.e., Eqs.~\eqref{eq:genperiodic1} and~\eqref{eq:genperiodic2}.
%
%
\subsection{Micromorphic Computational Homogenization}
A micromorphic continuum theory is based on the introduction of additional kinematic variables capturing the microstructural displacement field, which include non-local effects on the macro level. The Cosserat continuum~\citep{Cosserat1909} is a micromorphic continuum where the additional kinematic variables take the microstructural rotations into account. In~\citep{Rokos2018}, the micromorphic theory is used to construct a homogenization framework tailored to cellular elastomeric metamaterials. Here, the characteristic deformation mode, corresponding to the pattern transformation, is extracted from the displacement field, and an additional variable controlling its magnitude is introduced.

The following decomposition ansatz of the kinematic field then holds
\begin{equation}
\vec{u}(\vec{X},\vec{\zeta}) = \vec{v_{0}}(\vec{X}) + \sum^{n}_{i = 1} v_{i}(\vec{X})\vec{\varphi_{i}}(\vec{X},\vec{\zeta}) + \vec{w}(\vec{X},\vec{\zeta}),
\label{eq:displacementdecomposition}
\end{equation}
in which the vector function~$\vec{v}_{0}(\vec{X})$ corresponds to the mean effective displacement field, $v_{i}(\vec{X})$ are scalar fields representing the magnitudes of the predetermined long-range correlated patterning modes~$\vec{\varphi}_{i}(\vec{X},\vec{\zeta})$, and~$\vec{w}(\vec{X},\vec{\zeta})$ is the remaining microfluctuation field. The notion of ensemble averaging is adopted, recall Eq.~\eqref{eq:EnsembleAverage} and the discussion therein, which is reflected in Eq.~\eqref{eq:displacementdecomposition} through the dependence of the micro-fluctuating quantities on the microstructural translation~$\vec{\zeta} \in Q$. For the problem considered here, only a single spatially correlated mode is considered (i.e., $n = 1$, shown in Fig.~\ref{fig:PatTrans_b}), which is considered as an input quantity (prior kinematic knowledge), and can be obtained through a Bloch-type analysis~\citep{Bertoldi2008}, estimated analytically from full-scale numerical simulations~\citep{Rokos2018}, or identified experimentally.

To establish a computational homogenization scheme, a Taylor expansion of the coarse fields~$\vec{v}_0$ and~$v_1$ is performed in analogy to the first- and second-order computational methods (cf. Eqs.~\eqref{eq:TaylorExpansionFirst} and~\eqref{eq:TaylorExpansion}), yielding
\begin{equation}
\vec{x}_{\mathrm{m}} =  \bs{F}\cdot\vec{X}_\mathrm{m} + \big[v_{1} + \vec{X}_\mathrm{m} \cdot \vec{\nabla}v_{1}\big]\vec{\varphi}_1(\vec{X}_\mathrm{m}) + \vec{w}(\vec{X}_\mathrm{m}),
\label{eq:micromorphic:u}
\end{equation}
where the translation variable~$\vec{\zeta} \in Q$ has been replaced by the spatial variable~$\vec{X}_\mathrm{m} \in \Omega_\mathrm{m}$, effectively replacing the expensive ensemble averaging by substantially cheaper volume averaging. Note that since the patterning mode~$\vec{\varphi}_1$ is defined with respect to the reference microscopic configuration~$\vec{X}_\mathrm{m}$, macroscopic rotation needs to be factored out especially in cases in which it is non-negligible. This can be achieved through polar decomposition of the macroscopic deformation gradient~$\boldsymbol{F} = \boldsymbol{R} \cdot \boldsymbol{U}$, where~$\boldsymbol{R}$ is the macroscopic rotation tensor and~$\boldsymbol{U}$ the macroscopic stretch tensor. This effectively means that the deformation gradient~$\boldsymbol{F}$ is replaced with~$\boldsymbol{U}$ in Eq.~\eqref{eq:micromorphic:u}, see also~\cite[][Section~2.1]{Bree2019}. Unlike the second-order method, the non-local information of the micromorphic model is contained in the magnitude of the patterning mode~$v_1$ and its spatial variation~$\vec{\nabla}v_1$. For more details see also~\citep[][Section~5.1]{Rokos2018}. The kinematic approximation of Eq.~\eqref{eq:micromorphic:u} is considered over a local neighbourhood of a macroscopic point~$\vec{X}$, spanned by the adopted RVE, over which the micro-fluctuation field~$\vec{w}$ can be computed. Inserting the decomposition of Eq.~\eqref{eq:micromorphic:u} into the internal virtual work, two macroscopic balance equations result,
\begin{align}
\vec{\nabla}\cdot\boldsymbol{\Theta}^{\mathrm{T}} &= \vec{0}, \quad \vec{X} \in \Omega, \label{eq:micromorphic:macroproblem_a}\\
\vec{\nabla} \cdot \vec{\Lambda} - \Pi &= 0, \quad \vec{X} \in \Omega, \label{eq:micromorphic:macroproblem_b}
\end{align}
neglecting, again, any body or inertia forces, along with the following homogenized stress quantities
\begingroup
\allowdisplaybreaks
\begin{align}
\boldsymbol{\Theta}(\vec{X}) &= \frac{1}{|\Omega_{\mathrm{m}}|}\int_{\Omega_{\mathrm{m}}}\boldsymbol{P}_\mathrm{m}\,\mathrm{d}\Omega_{\mathrm{m}},\\
\Pi(\vec{X}) &= \frac{1}{|\Omega_{\mathrm{m}}|}\int_{\Omega_{\mathrm{m}}}\boldsymbol{P}_\mathrm{m}:\vec{\nabla}_{\mathrm{m}}\vec{\varphi}_{1}\,\mathrm{d}\Omega_{\mathrm{m}},\\
\vec{\Lambda}(\vec{X}) &= \frac{1}{|\Omega_{\mathrm{m}}|}\int_{\Omega_{\mathrm{m}}}\boldsymbol{P}_{\mathrm{m}}^{\mathrm{T}}\cdot\vec{\varphi}_{1} + \vec{X}_{\mathrm{m}}[\boldsymbol{P}_{\mathrm{m}}:\vec{\nabla}_{\mathrm{m}}\vec{\varphi}_{1}]\,\mathrm{d}\Omega_{\mathrm{m}}.
\end{align}
\endgroup

At the microscale the classical balance law of Eq.~\eqref{eq:balancemicro} applies, which requires in addition to periodicity of~$\vec{w}$ in Eq.~\eqref{eq:pbcs} also orthogonality of~$\vec{w}$ with respect to a constant field, fluctuation mode~$\vec{\varphi}_1$, and linear component~$\vec{\varphi}_1\vec{X}_\mathrm{m}$, to guarantee uniqueness of the solution. Unlike the second-order homogenization scheme, the four RVE corner points are free, and rigid body motions are restricted by a combination of periodicity conditions and the orthogonality with respect to a constant field. For more details on implementation and theory see~\cite{Bree2019} and~\cite{Rokos2019a}.
%
%
\subsection{Concluding Remarks}
Each of the presented homogenization schemes relies on a classical continuum at the microscale level. The first-order method employs a classical continuum description also at the macro-scale level, while the second-order method extends the balance equation by including higher-order stresses arising from strain gradients. The micromorphic method, on the contrary, introduces an additional scalar balance law governing the macroscopic magnitude of the fluctuation pattern. Whereas the first-order and second-order methods require no prior information on the considered kinematics, the micromorphic scheme assumes the typical long-range correlated patterning field to be known in advance for the adopted kinematic ansatz.
%
%
\section{Numerical Examples and Comparison}
\label{sec:Results}
This section provides a performance analysis of the homogenization methodologies discussed in Section~\ref{sec:Methodology}, applied to pattern transforming elastomeric metamaterials. The accuracy of the enriched homogenization solutions is assessed relative to the reference ensemble-averaged solution, whereas the first-order homogenization solution is considered as the current standard (most widely used). Three distinct loading cases are evaluated, comparing mechanical responses obtained by each of the methods. To this end, first an infinitely wide specimen loaded in compression is considered. Second, a specimen subjected to bending is evaluated. A finite rectangular specimen, subjected to a compressive load with two free boundaries experiencing an auxetic effect, is analysed as the final example.
%
%
\subsection{Uniform Compression}
\label{sect:inifinte_compression}
\begin{figure}
    \centering
    \includegraphics{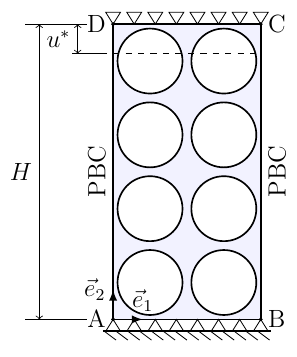}
    \caption{Uniform compression case. A sketch of the considered simulation domain and boundary conditions for a specimen with scale ratio~$H/l = 4$. Periodic boundary conditions are enforced along the two vertical edges AD and BC, representing an infinitely wide structure.}
    \label{fig:BCInfiniteCompression}
\end{figure}

\begin{figure}
	\centering
	\includegraphics{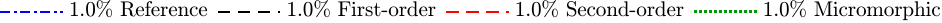}\\
	\includegraphics{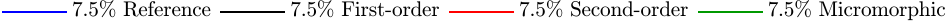}\\
	\subfloat[vertical displacement]{\includegraphics{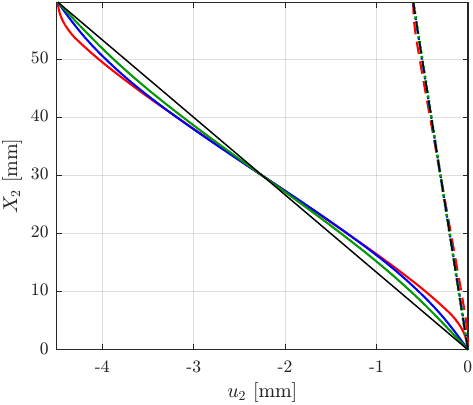}\label{fig:InComDeformationSR6_a}}
	\hspace{0.3cm}
	\subfloat[vertical deformation]{\includegraphics{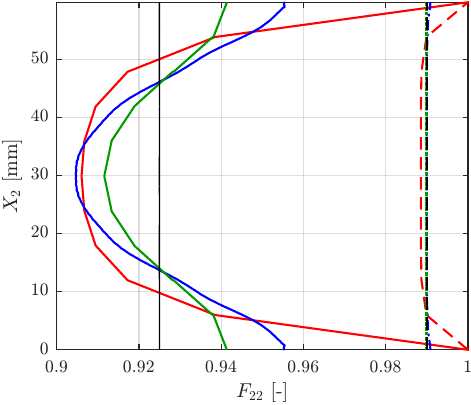}\label{fig:InComDeformationSR6_b}}
	\caption{Uniform compression case. Comparison of~(a) vertical displacement component~$u_2(X_2)$, and~(b) deformation gradient~$F_{22}(X_2)$ over the specimen's height. Overall applied compressive strain corresponds to~$1\%$ and~$7.5\%$, for a specimen with scale ratio~$H/l = 6$.}
	\label{fig:InComDeformationSR6}
\end{figure}

The first example considers an infinite layer of elastomeric material loaded in compression along the vertical direction. The specimen's height is~$H = n_{v}l$, where~$n_{v}$ is the number of unit cells in the vertical direction. The simulation domain spans the region~$2l \times H$, as shown in Fig.~\ref{fig:BCInfiniteCompression}, whereby only two unit cells are modelled in the horizontal direction, which is sufficient to capture the complete pattern transformation. The infinite structure is mimicked through periodicity along the two vertical edges, AD and BC. The bottom edge AB is fixed, whereas the top edge CD is displaced downwards by~$u^{*}$. As a result of these kinematic constraints, the macroscopic problem can be modelled as a one-dimensional continuum along~$\vec{e}_2$, with the following essential boundary conditions. For the first-order computational homogenization:
\begin{equation}
u_{2}(0) = 0,\, u_{2}(H) = u^{*},
\end{equation}
for the second-order method:
\begin{equation}
u_{2}(0) = 0,\, u_{2}(H) = u^{*},\quad F_{22}(0) = F_{22}(H) = 1,
\label{eq:infinite:dbc}
\end{equation}
and for the micromorphic approach:
\begin{equation}
v_{0,2}(0) = 0,\, v_{0,2}(H) = u^{*},\quad v_{1}(0) = v_{1}(H) = 0.
\end{equation}
Note that the choice of~$F_{22} = 1$ in Eq.~\eqref{eq:infinite:dbc} is motivated by physical insights of no deformation at the top and bottom part of the domain boundary. Although not presented for conciseness, numerical investigations showed high influence of prescribed values of~$F_{22}$ on achieved accuracy in comparison with DNS. For all methods, the macroscopic displacement fields are discretized by piece-wise quadratic one-dimensional elements with a two-point Gauss integration rule. For the second-order method, the Lagrange multiplier field and relaxed deformation gradient field are discretized using piece-wise linear shape functions. In one-dimensional setting considered here, as an alternative to the mixed formulation, Hermite polynomial elements could be used for the second-order scheme, which are not considered for consistency with the other two-dimensional examples discussed below. For discretization, domains with scale ratios~$H/l \leq 10$ were calculated using fine meshes of~$10$ one-dimensional macroscopic elements to ensure sufficiently high kinematic freedom to accurately capture boundary layers spanning the entire domain (see ahead Fig.~\ref{fig:InComDeformationSR6_b}), whereas simulations at higher scale ratios~$H/l > 10$ have as many elements as unit cells due to the lower complexity of the resulting kinematic fields (forming a distinct plateau in~$F_{22}$ component close to the mid-height of the specimen; not shown herein, see, e.g., \citealt{Ameen2018b}).
\begin{figure}
	\centering
	\begin{tabular}{@{}ll}
	\subfloat[DNS]{\includegraphics[scale=1.05]{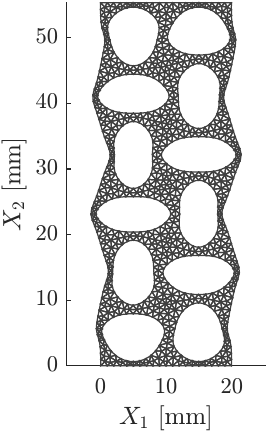}\label{fig:InComRVEs_a}} &
	\subfloat[first-order]{\includegraphics{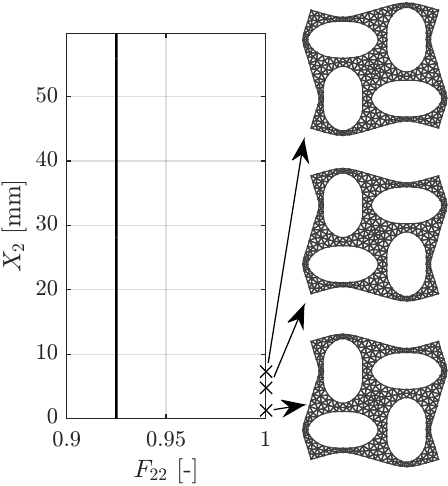}\label{fig:InComRVEs_b}}\\
	\subfloat[second-order]{\includegraphics{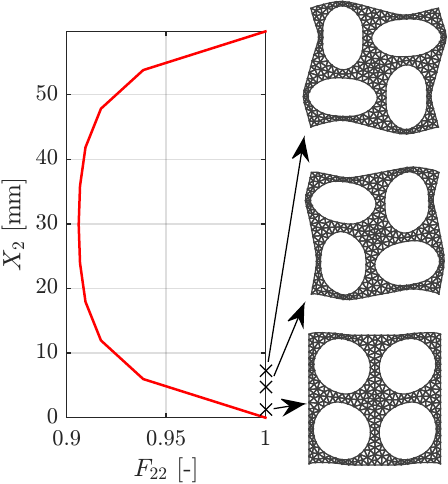}\label{fig:InComRVEs_c}} &
	\subfloat[micromorphic]{\includegraphics{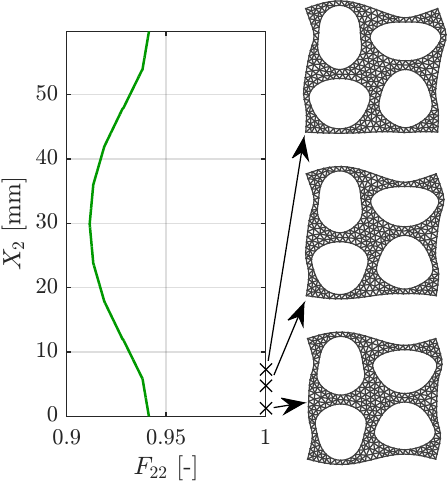}\label{fig:InComRVEs_d}}
	\end{tabular}
	\caption{Uniform compression case. Deformed configuration of an infinitely wide specimen with scale ratio~$H/l = 6$, subjected to~$7.5\%$ nominal compressive strain. (a)~DNS solution corresponding to zero microstructural translation~$\vec{\zeta} =\vec{0}$. $F_{22}$ component of the macroscopic deformation gradient plotted against the vertical coordinate in the reference configuration, including deformed RVEs of the first three integration points, corresponding to the~(b) first-order, (c) second-order, and~(d) micromorphic computational homogenization schemes.}
	\label{fig:InComRVEs}
\end{figure}
\begin{figure}
	\centering
	\includegraphics{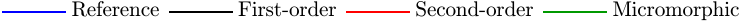}\\
	\subfloat[$H/l = 6$]{\includegraphics{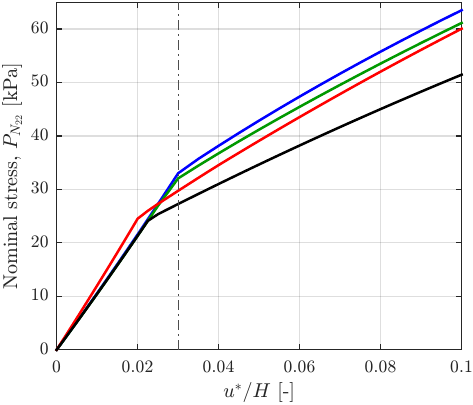}\label{fig:StressStrainInCom_a}}
	\hspace{0.3cm}
	\subfloat[$H/l = 12$]{\includegraphics{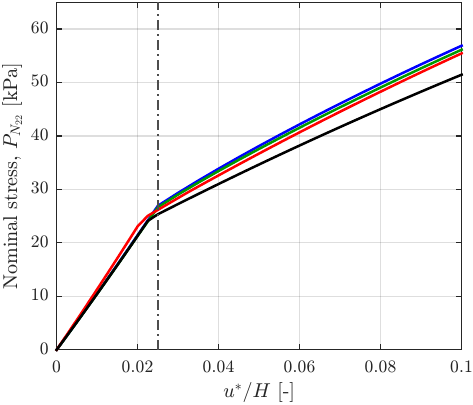}\label{fig:StressStrainInCom_b}}
	\caption{Uniform compression case. Stress--strain diagram for scale ratios (a)~$H/l = 6$ and (b)~$H/l = 12$, obtained via DNS and the first-order, second-order, and micromorphic homogenization schemes.}
	\label{fig:StressStrainInCom}
\end{figure}
\begin{figure}
	\centering
	\includegraphics{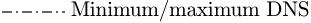}\\
	\includegraphics{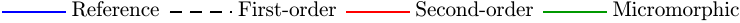}\\
	\subfloat[$u^{*}/H = 0.01$]{\includegraphics{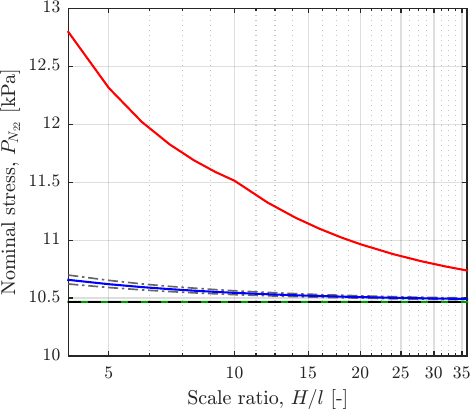}\label{fig:ScaleRatioInCom_a}}
	\hspace{0.3cm}
	\subfloat[$u^{*}/H = 0.075$]{\includegraphics{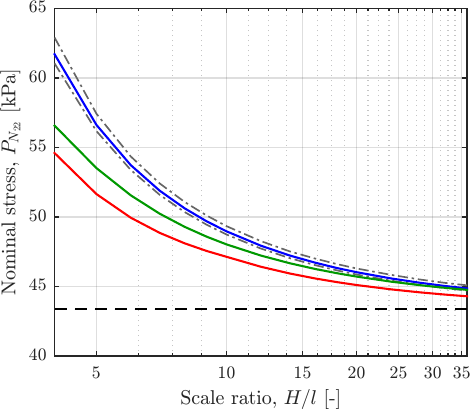}\label{fig:ScaleRatioInCom_b}}
	\caption{Uniform compression case. Nominal compressive stress in the~(a) pre-bifurcation and~(b) post bifurcation regime as a function of the scale ratio~$ H/l \in [ 4 , 36 ] $.}
	\label{fig:ScaleRatioInCom}
\end{figure}
\begin{figure}
	\centering
	\includegraphics{MinMaxDNSLegend.pdf}\\
	\includegraphics{ScaleSepLegend.pdf}\\
	\subfloat[$u^{*}/H = 0.01$]{\includegraphics{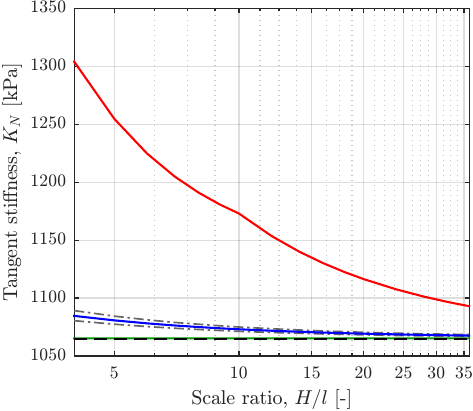}\label{fig:ScaleRatioInComStiffness_a}}
	\hspace{0.3cm}
	\subfloat[$u^{*}/H = 0.075$]{\includegraphics{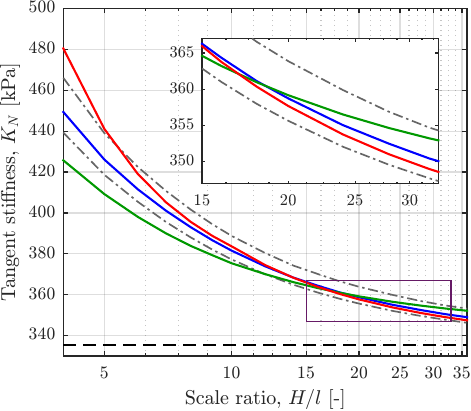}\label{fig:ScaleRatioInComStiffness_b}}
	\caption{Uniform compression example. Tangent stiffness in the~(a) pre-bifurcation and~(b) post-bifurcation regime as a function of the scale ratio~$ H/l \in [ 4 , 36 ] $.}
	\label{fig:ScaleRatioInComStiffness}
\end{figure}

The macroscopic kinematic quantities in the pre- and post-bifurcation regimes for a fixed scale ratio~$ H/l = 6 $ are compared in Fig.~\ref{fig:InComDeformationSR6}. Here, the reference solution indicates that during pre-bifurcation, no strain gradients are present, which is accurately captured by the micromorphic and first-order method. The second-order scheme exhibits, however, artificial gradient effects close to the two horizontal boundaries. Post-bifurcation, the micromorphic method shows a better deformation trend compared to the second-order scheme, which again exhibits too strong gradient effects, which are mainly caused by the kinematic constraints on~$F_{22}$ in Eq.~\eqref{eq:infinite:dbc}. Note that~$F_{22}(0) = F_{22}(H) \approx 0.95$ in Fig.~\ref{fig:InComDeformationSR6_b} for the reference solution, implying that the kinematic constraints in the second-order scheme are too restrictive. However, relaxing these boundary conditions introduces large errors in the post-bifurcation regime, resulting in an overall response that is too compliant.

Fig.~\ref{fig:InComRVEs} presents a deformed DNS configuration along with three RVEs positioned close to the specimen's bottom edge for each homogenization method. The gradient effect on the deformation of the individual RVEs is clearly visible close to the two boundaries. The first-order method (Fig.~\ref{fig:InComRVEs_b}) does not include strain gradients and provides the same compressed RVE pattern at each integration point, even close to the boundary (in fact, a single RVE could have been used to perform this simulation). The RVE deformations obtained by the second-order method (Fig.~\ref{fig:InComRVEs_c}), show gradual compression in the vertical direction of individual RVEs with growing distance from the fixed boundary, with little gradient of the patterning field along the vertical direction. The micromorphic RVEs, shown in Fig.~\ref{fig:InComRVEs_d}, show a slightly stronger gradient of the patterning. In both cases, nevertheless, the pattern is not fully representative for what can be observed in the DNS result.

The obtained nominal stress--strain diagrams are provided in Fig.~\ref{fig:StressStrainInCom} for scale ratios~$6$ and~$12$. The reference solution shows that the bifurcation strain (indicated by the black dashed vertical line) decreases with increasing scale ratio, and that the bifurcation strain is captured accurately by the micromorphic homogenization method (the error in bifurcation stress is approximately~$2.8\%$ and~$0.82\%$ for scale ratios~$6$ and~$12$). The first-order method, which is insensitive to the macroscopic size, provides a bifurcation strain~$u^{*}/H \approx 0.0225$. Interestingly, the second-order method exhibits an opposite effect, in which the bifurcation strain increases with increasing scale ratio. The final buckling strain corresponds to~$u^{*}/H \approx 0.0225$, which is expected since the first-order solution should be recovered for~$H/l \rightarrow \infty$.

The effect of the macroscopic size on the stress response is shown in Fig.~\ref{fig:ScaleRatioInCom}, where the homogenized stresses at fixed nominal strains~$u^{*}/H = 0.01$ and $u^{*}/H = 0.075$ are plotted against the scale ratio~$ H/l \in [ 4, 36 ] $. Before the pattern transformation occurs, no non-local effects (and thus almost no size effects) can be observed in the reference, first-order, and micromorphic homogenized solutions (Fig.~\ref{fig:ScaleRatioInCom_a}). This is not the case, however, for the second-order homogenization scheme, which shows an excessively stiff response in the pre-bifurcation regime due to the kinematic constraint~$F_{22} = 1$. In the post-bifurcation regime, the second-order and micromorphic homogenization schemes capture the overall trend, but both underestimate the effective stress. For the smallest scale ratio considered, i.e., $H/l = 4$, the first-order homogenization entails approximately~$30\%$, second-order~$11.5\%$, and micromorphic~$7.5\%$ relative error.

The effective stiffnesses, considered again at~$u^{*}/H = 0.01$ and~$u^{*}/H = 0.075$ of the overall nominal strain and expressed as a function of the scale ratio~$H/l,$ are shown in Fig.~\ref{fig:ScaleRatioInComStiffness}. Pre-bifurcation, similar trends as for the nominal stresses are observed (cf. Fig.~\ref{fig:ScaleRatioInCom_a}), whereas post-bifurcation (Fig.~\ref{fig:ScaleRatioInComStiffness_b}), observed differences are approximately equal for both non-local schemes. For scale ratios~$H/l > 10$, nevertheless, the second-order method follows the reference trend more accurately compared to the micromorphic method.
%
%
\subsection{Uniform Bending}
\label{sect:uniform_bending}
\begin{figure}
    \centering
    \includegraphics{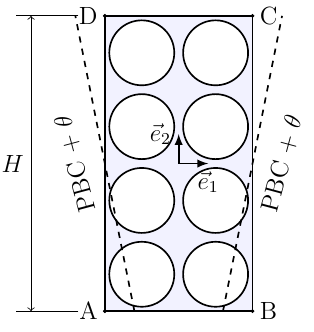}
    \caption{Uniform bending case. A sketch of the simulation domain and boundary conditions for a specimen with scale ratio~$H/l = 4$. Superposed periodic boundary conditions are enforced along the vertical edges AD and BC to mimic an infinitely wide structure.}
    \label{fig:BC_InBend}
\end{figure}

Similarly to the previous example, a semi-infinite domain is considered, which is subjected to uniform bending. The simulation domain again spans a region~$2l \times H$, where an integer number of unit cells along the vertical direction is considered, cf. Fig.~\ref{fig:BC_InBend}. The two vertical edges, AD and BC, deform periodically on top of the uniform bending contribution. The coordinate system is shifted to the midpoint of the simulation domain such that the neutral axis corresponds to~$X_2 = 0$. The neutral axis is furthermore fixed in the vertical direction by constraining horizontal and vertical displacements at the centre left and right nodes, whereas the top and bottom edges are considered as free surfaces. Mathematically, the applied boundary conditions are expressed as follows. For the first-order method:
\begin{equation}
\vec{u}(\Gamma_\mathrm{AD}) = \vec{u}(\Gamma_\mathrm{BC}) + \theta X_2 \vec{e}_1,
\end{equation}
for the second-order method:
\begin{equation}
\vec{u}(\Gamma_\mathrm{AD}) = \vec{u}(\Gamma_\mathrm{BC}) + \theta X_2 \vec{e}_1,\quad F_{11}(\Gamma_\mathrm{AD}) = F_{11}(\Gamma_\mathrm{BC}),\, F_{22}(\Gamma_\mathrm{AD}) = F_{22}(\Gamma_\mathrm{BC}),
\end{equation}
and for the micromorphic method:
\begin{equation}
\vec{v}_{0}(\Gamma_\mathrm{AD}) = \vec{v}_{0}(\Gamma_\mathrm{BC}) + \theta X_2 \vec{e}_1,\quad v_{1}(\Gamma_\mathrm{AD}) = v_{1}(\Gamma_\mathrm{BC}).
\end{equation}
The bending angle~$\theta$ is prescribed such that the nominal strain at the top and bottom edges (expressed as~$\theta H/(4l)$) does not exceed~$15\%$. The macroscopic discretization consists of $8$-node two-dimensional square quadratic quadrilateral elements with edge size~$l$, integrated by a~$ 2 \times 2 $ Gauss integration rule. The relaxed deformation gradient field~$\widehat{\bs{F}}$, present in the second-order homogenization method, is discretized by $4$-node bi-linear quadrilaterals. More details and performance analysis of various two-dimensional element types for micromorphic scheme can be found in~\cite{Rokos2020}, and for second-order computational homogenization, e.g., in~\cite{Kouznetsova2004}.
\begin{figure}
    \centering
   	\begin{tabular}{@{}ll}
    \subfloat[DNS]{\raisebox{0.5cm}{\includegraphics[scale = 0.95]{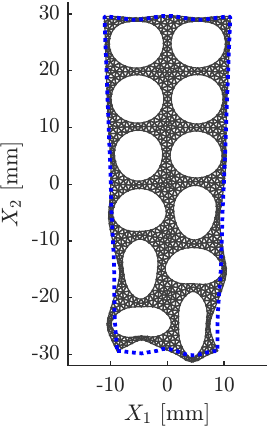} \label{fig:RVEsInBend_a}}} &
    \subfloat[first-order]{\includegraphics[scale = 0.95]{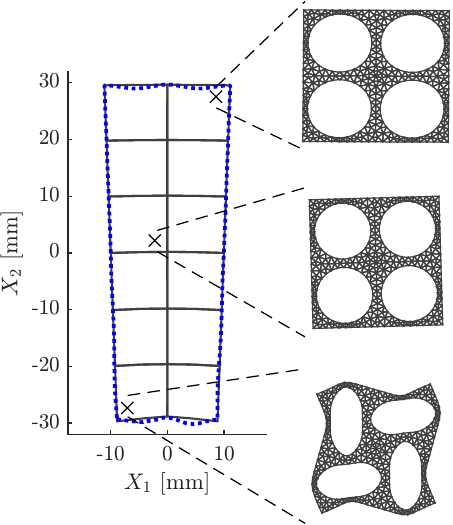}\label{fig:RVEsInBend_b}}\\
    \subfloat[second-order]{\includegraphics[scale = 0.95]{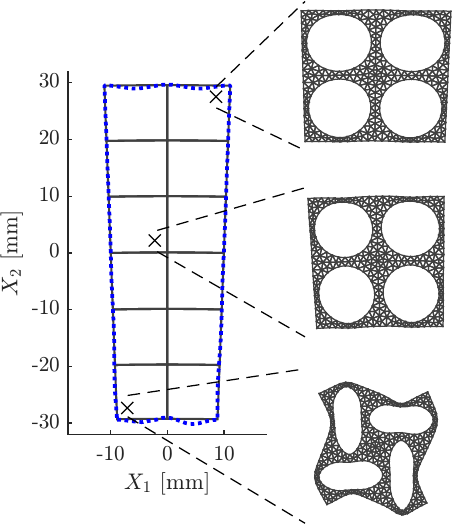} \label{fig:RVEsInBend_c}} &
    \subfloat[micromorphic]{\includegraphics[scale = 0.95]{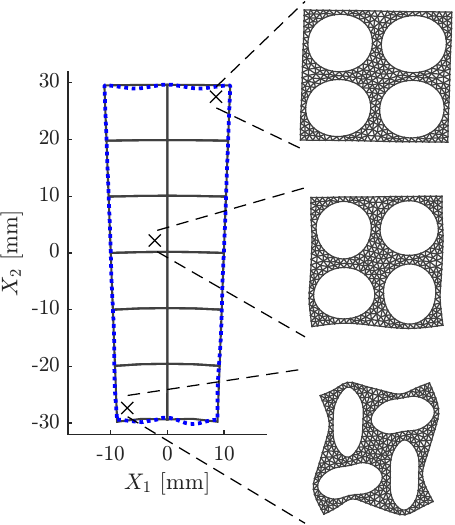}\label{fig:RVEsInBend_d}}
	\end{tabular}
    \caption{Uniform bending case. An infinitely wide specimen with scale ratio~$6$ bent up to~$\theta H/(4l) = 0.1125$. (a) DNS solution with zero microstructural shift~$\vec{\zeta} =\vec{0}$. The macroscopic deformed structures including deformed RVEs corresponding to three macroscopic integration points for the~(b) first-order, (c) second-order, and~(d) micromorphic homogenization method. The dotted blue line indicates the deformed contour of the ensemble-averaged solution.}
    \label{fig:RVEsInBend}
\end{figure}
\begin{figure}
    \centering
    \includegraphics{ScaleSepLegend.pdf}\\
    \subfloat[stress--strain diagram]{\includegraphics{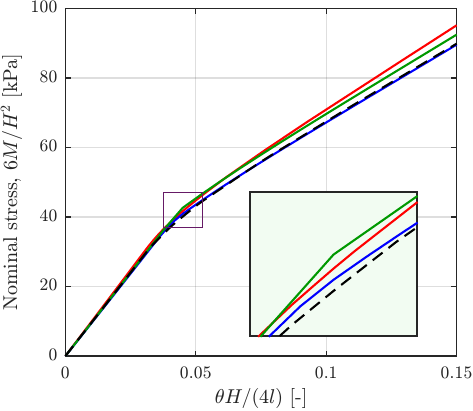}\label{fig:InBendSR12_a}}
    \hspace{0.2cm}
    \subfloat[homogenized stress~$P_{11}$]{\includegraphics{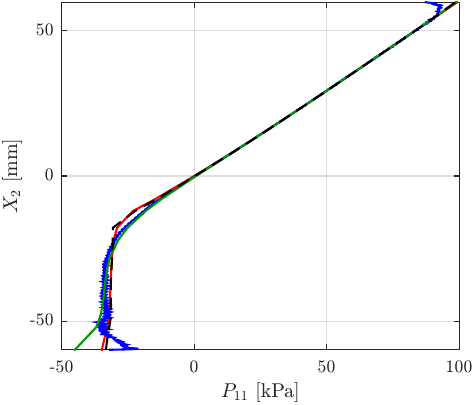}\label{fig:InBendSR12_b}}
    \caption{Uniform bending case corresponding to scale ratio~$12$. (a) Stress--strain diagram; a close-up of the bifurcation point in the inset. (b) $P_{11}$ stress component along the vertical reference coordinate~$X_2$ at~$\theta H/(4l) = 0.1125$.}
    \label{fig:InBendSR12}
\end{figure}
\begin{figure}
    \centering
    \includegraphics{MinMaxDNSLegend.pdf}\\
    \includegraphics{ScaleSepLegend.pdf}\\
    \subfloat[$\theta H/(4l) = 0.015$]{\includegraphics{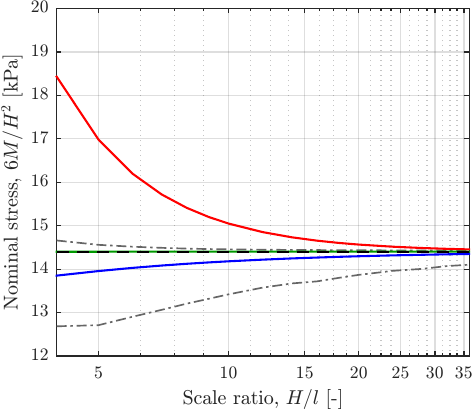}\label{fig:ScaleSepInBend_a}}
    \hspace{0.3cm}
    \subfloat[$\theta H/(4l) = 0.1125$]{\includegraphics{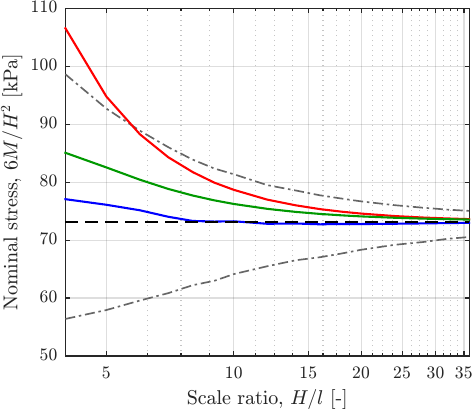}\label{fig:ScaleSepInBend_b}}
    \caption{Uniform bending case. Nominal stress in the~(a) pre-bifurcation and~(b) post-bifurcation regime as a function of the scale ratio~$ H/l \in [ 4 , 36 ] $.}
    \label{fig:ScaleSepInBend}
\end{figure}

The deformed configurations, obtained for all methods, are shown in Fig.~\ref{fig:RVEsInBend} for an applied nominal strain~$\theta H/(4l) = 0.1125$ (i.e., in the post-bifurcation regime). Since only the bottom half of the specimen is subjected to compression, the pattern transformation is limited to the region situated below the neutral axis, cf. Fig.~\ref{fig:RVEsInBend_a}. The macroscopic deformation along with deformed RVEs corresponding to three macroscopic integration points are shown in Figs.~\ref{fig:RVEsInBend_b}--\ref{fig:RVEsInBend_d}. Here a clear distinction in the RVE deformations for the different methods emerges. In particular, for the first-order homogenization scheme, individual RVEs have parallelepiped shape and show uniform buckling pattern of individual holes (Fig.~\ref{fig:RVEsInBend_b}). For the second-order homogenization method, the bending effect of the higher-order gradient~$\partial F_{11}/\partial X_{2} = G_{211} \neq 0$ applied to individual RVEs can clearly be observed, resulting in trapezoidal-shaped RVEs (Fig.~\ref{fig:RVEsInBend_c}). Finally, the effect of the pattern gradient~$\partial v_1/\partial X_2$ is present in the micromorphic scheme, resulting in parallelepiped-shaped RVEs with non-uniform buckling pattern of holes along the vertical direction (Fig.~\ref{fig:RVEsInBend_d}).

The nominal bending stress, defined as~$6M/H^2$, where~$M$ denotes the moment acting on the vertical specimen edge, is plotted as a function of the nominal strain~$\theta H/(4l)$ for a scale ratio~$12$ in Fig.~\ref{fig:InBendSR12_a}. Even though both enriched homogenization schemes provide similar stress--strain results, they systematically overestimate the stress in the post-bifurcation regime. The micromorphic method exhibits a clear kink when moving from pre- to post-bifurcation regime, whereas the first- and second-order methods show a more gradual bifurcation transition. The corresponding homogenized stress component of the first Piola--Kirchhoff stress tensor, $P_{11}(X_2)$, is shown in Fig.~\ref{fig:InBendSR12_b} as a function of the vertical coordinate~$X_2$ for scale ratio~$12$ and a nominal strain~$\theta H/(4l) = 0.1125$. Due to the pattern transformation, the bottom half of the specimen displays a distinct plateau, whereas the top part shows a linear profile. All homogenization methods considered provide relatively accurate predictions.

The scale separation plots corresponding to strains~$\theta H/(4l) = 0.015$ and~$\theta H/(4l) = 0.1125$, given in Fig.~\ref{fig:ScaleSepInBend}, reflect the scale ratio effect in the pre- and post-bifurcation regime. In both cases, the micromorphic prediction is closer to the reference solution compared to the second-order method. The size effect is, however, not as strong as for the previously discussed compression case, through which the first-order method (neglecting size effects) provides the most accurate solution in the pre- and post-bifurcation regimes. Note that although the ensemble averaged solution is not varying much with the scale ratio, individual microstructural translations do result in a large variance (cf. Fig.~\ref{fig:ScaleSepInBend}).
%
%
\subsection{Compression of a Finite Specimen}
\label{sect:finte_compression}
\begin{figure}
    \centering
    \includegraphics{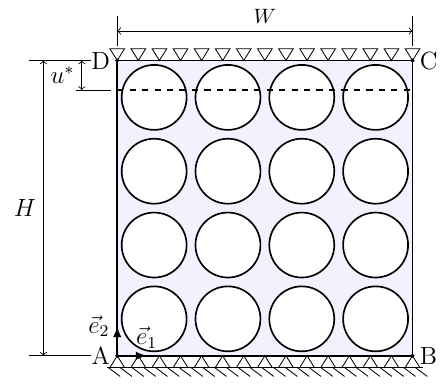}
    \caption{Finite specimen compression case. Illustration of the simulation~$ 4l \times 4l $ domain subjected to a vertical compressive load. The two lateral edges, AD and BC, are free surfaces.}
    \label{fig:BC_FiCom}
\end{figure}

The final case considers a finite rectangular domain spanning~$n\times n$ unit cells subjected to compression, cf. Fig.~\ref{fig:BC_FiCom}. The applied compression acts along the vertical direction, and is induced by fixing the bottom horizontal edge AB while  displacing the top horizontal edge CD downwards by~$u^{*}$. Unlike the uniform compression case of Section~\ref{sect:inifinte_compression}, the two vertical edges AD and BC are now free surfaces, allowing for compliant boundary layers and auxetic effects. Mathematically, the applied boundary conditions are expressed for the first-order homogenization method as:
\begin{equation}
\vec{u}(\Gamma_\mathrm{AB}) = \vec{0},\, \vec{u}(\Gamma_\mathrm{CD}) = -u^{*}\, \vec{e}_{2},
\end{equation}
for the second-order method:
\begin{equation}
\vec{u}(\Gamma_\mathrm{AB}) = \vec{0},\, \vec{u}(\Gamma_\mathrm{CD}) = -u^{*}\, \vec{e}_{2},\quad F_{22}(\Gamma_\mathrm{AB}) = F_{22}(\Gamma_\mathrm{CD}) = 1,
\label{eq:bc:ch2}
\end{equation}
and for the micromorphic method:
\begin{equation}
\vec{v}_{0}(\Gamma_\mathrm{AB}) = \vec{0}, \, \vec{v}_{0}(\Gamma_\mathrm{CD}) = -u^{*}\, \vec{e}_{2}, \quad v_{1}(\Gamma_\mathrm{AB}) = v_{1}(\Gamma_\mathrm{CD}) = 0.
\end{equation}
In order to avoid global buckling, which has been addressed elsewhere~\citep[cf.][]{Bree2019,Wu2023}, and to focus on boundary layers and non-local behaviour, only one half of the specimen is modelled, introducing a vertical symmetry axis at the middle of the specimen along which the horizontal displacements are restricted while the vertical displacements are left free. The same discretization and integration rules as those in Section~\ref{sect:uniform_bending} are used.
\begin{figure}
    \centering
   	\begin{tabular}{@{}ll}
    \subfloat[DNS]{\raisebox{0.8cm}{\includegraphics[scale = 0.95]{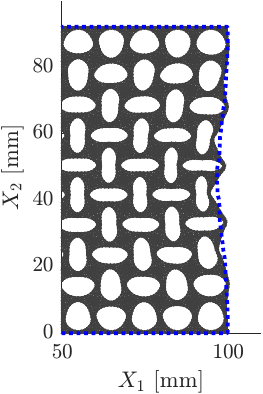} \label{fig:RVEsFiCom_a}}} &
    \subfloat[first-order]{\includegraphics[scale = 0.95]{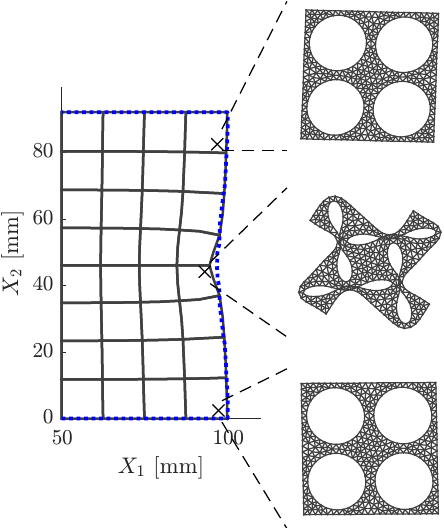} \label{fig:RVEsFiCom_b}} \\
    \subfloat[second-order]{\includegraphics[scale = 0.95]{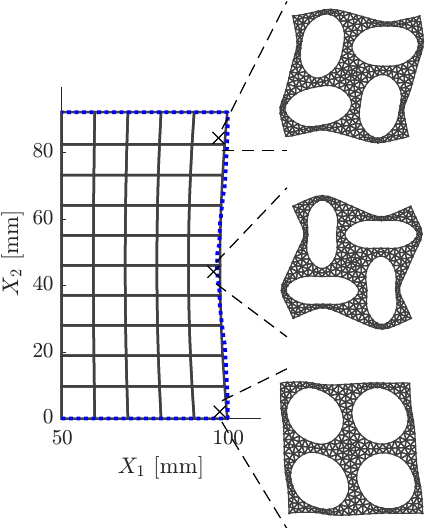} \label{fig:RVEsFiCom_c}} &
    \subfloat[micromorphic]{\includegraphics[scale = 0.95]{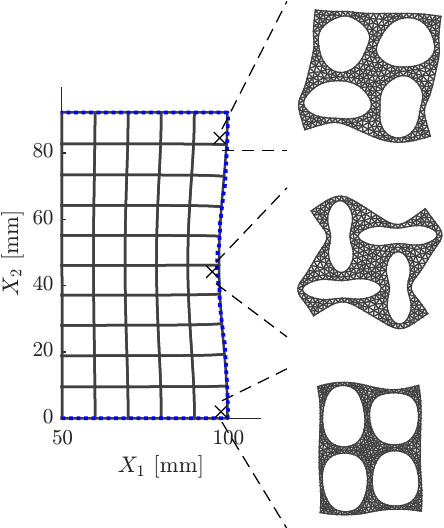} \label{fig:RVEsFiCom_d}}
    \end{tabular}
    \caption{Finite specimen compression case. Right half of a deformed~$10l \times 10l$ specimen obtained by (a)~DNS subjected to~$u^{*}/H = 0.075$ compressive strain, corresponding to a zero microstructural translation~$\vec{\zeta} = \vec{0}$. The right half of the macroscopic deformed structure, including deformed RVEs positioned at three macroscopic integration points, corresponding to the~(b) first-order, (c)~second-order, and~(d) micromorphic homogenization method. The dotted blue line indicates the deformed contour of the ensemble-averaged solution.}
    \label{fig:RVEsFiCom}
\end{figure}
\begin{figure}
    \centering
    \includegraphics{StressLegend.pdf}\\
    \subfloat[$H/l = 4$]{\includegraphics{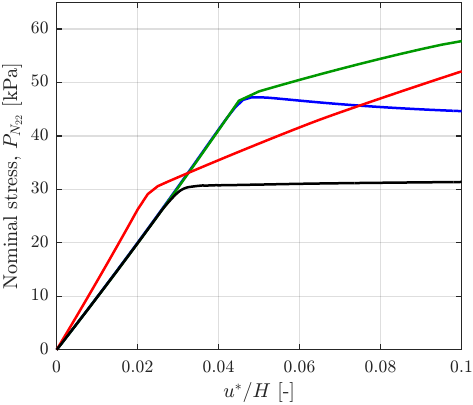}\label{fig:FiComStressStrain_a}}
    \hspace{0.3cm}
    \subfloat[$H/l = 10$]{\includegraphics{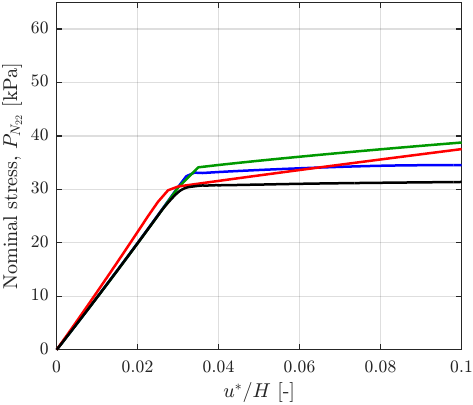}\label{fig:FiComStressStrain_b}}
    \caption{Finite specimen compression case. Stress--strain diagrams corresponding to scale ratios (a)~$4$, and (b)~$10$, obtained by the first-order, second-order, and micromorphic homogenization schemes, as compared to the reference ensemble averaged solution.}
    \label{fig:FiComStressStrain}
\end{figure}
\begin{figure}
    \centering
    \includegraphics{MinMaxDNSLegend.pdf}\\
    \includegraphics{ScaleSepLegend.pdf}\\
    \subfloat[$u^{*}/H = 0.01$]{\includegraphics[scale=0.833]{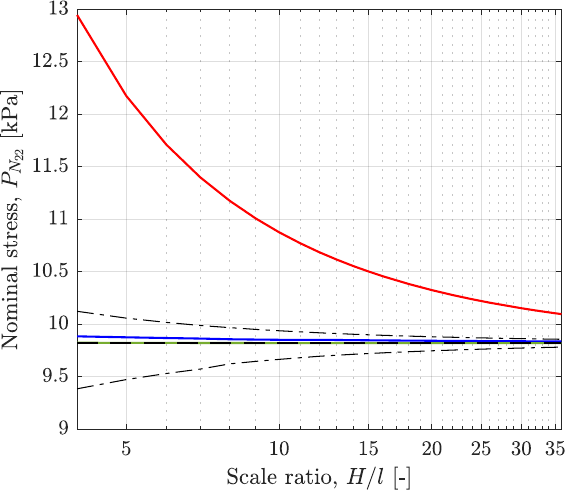}\label{fig:ScaleSepFiCom_a}}
    \hspace{0.3cm}
    \subfloat[$u^{*}/H = 0.05$]{\includegraphics[scale=0.833]{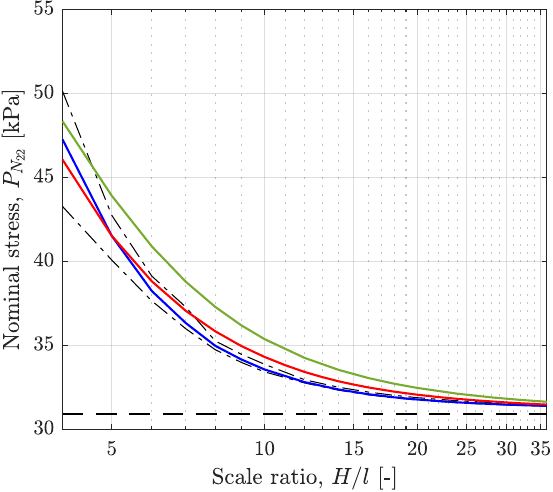}\label{fig:ScaleSepFiCom_b}}
    \caption{Finite specimen compression case. Nominal compressive stress in the~(a) pre-bifurcation and~(b) post-bifurcation regime as a function of the scale ratio~$ H/l \in [ 4 , 36 ] $.}
    \label{fig:ScaleSepFiCom}
\end{figure}
\begin{figure}
    \centering
    \includegraphics{MinMaxDNSLegend.pdf}\\
    \includegraphics{ScaleSepLegend.pdf}\\
    \subfloat[$u^{*}/H = 0.01$]{\includegraphics[scale=0.833]{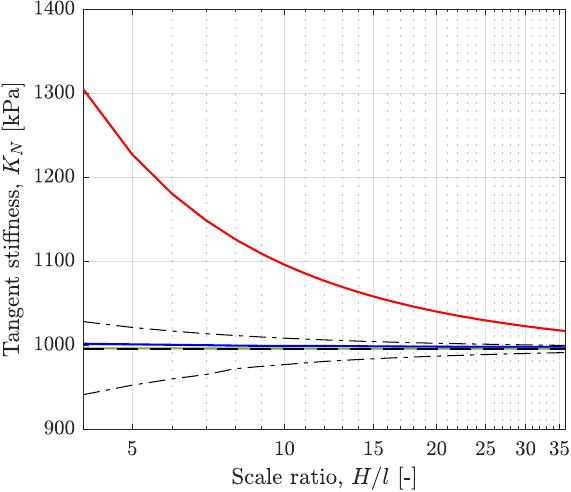}\label{fig:ScaleSepFiCom_stiffness_a}}
    \hspace{0.3cm}
    \subfloat[$u^{*}/H = 0.05$]{\includegraphics[scale=0.833]{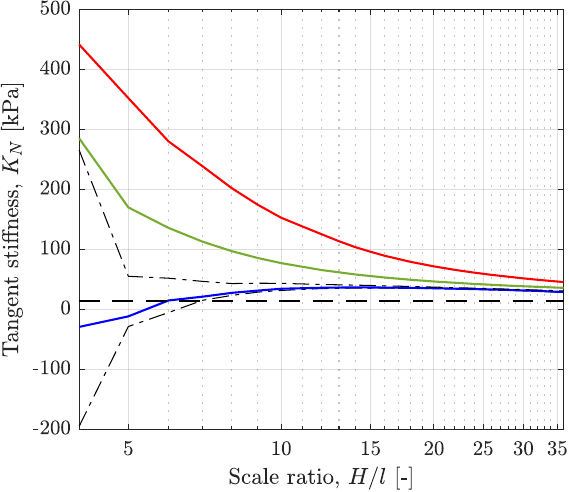}\label{fig:ScaleSepFiCom_stiffness_b}}
    \caption{Finite specimen compression case. Tangent stiffness in the~(a) pre-bifurcation and~(b) post-bifurcation regime as a function of the scale ratio~$ H/l \in [ 4 , 36 ] $.}
    \label{fig:ScaleSepFiCom_stiffness}
\end{figure}

Deformed configurations of a half specimen (due to symmetry), corresponding to scale ratio~$H/l = 10$ and subjected to~$7.5\%$ of overall compressive strain (i.e., post-bifurcation), are shown in Fig.~\ref{fig:RVEsFiCom}. The auxetic effect resulting from the pattern transformation is clearly visible and captured by all methods. The two enriched homogenization schemes (Figs.~\ref{fig:RVEsFiCom_c} and~\ref{fig:RVEsFiCom_d}) reproduce the auxetic behaviour more accurately than the first-order method (Fig.~\ref{fig:RVEsFiCom_b}), judging from the outer contours of the ensemble-averaged solution denoted by the blue dotted line. A significant drawback of the first-order method is the presence of a kink in the displacements at the centres of the two lateral edges. The overall shape of the compliant boundary and the deformed RVEs corresponding to the top and bottom Gauss integration points are most accurately captured by the micromorphic method. The investigated integration points close to the middle height of the specimen are similar for both enriched methods, whereas the RVEs corresponding to the first-order method deform severely close to the kinks.

The stress--strain diagrams, corresponding to~$4l \times 4l$ and~$10l \times 10l$ specimens, are provided in Fig.~\ref{fig:FiComStressStrain}. Similarly to the uniform compression case of Section~\ref{sect:inifinte_compression}, the bifurcation strain is not captured accurately by the second-order homogenization method, whereas it is captured adequately by the micromorphic scheme. This holds true especially for small scale ratios for which boundary effects play a significant role (clearly visible for~$H/l = 4$ in Fig.~\ref{fig:FiComStressStrain_a}), whereas for larger scale ratios the importance of non-local effects gradually diminishes and the achieved accuracy is of the same order of magnitude (as visible for~$H/l = 10$ in Fig.~\ref{fig:FiComStressStrain_b}). Ultimately, the common asymptote of the first-order scheme is reached for~$H/l \rightarrow \infty$, where non-local effects vanish completely. In the case of the~$4l \times 4l$ specimen, the post-bifurcation stiffness (i.e., the slope of the stress--strain diagram), although comparable for both enriched homogenization schemes, significantly overestimates the expected value, which even reveals softening (a negative slope). The observed softening results from the presence of the two vertical compliant boundaries, which for such a small scale ratio have a larger influence on the overall response than the stiff boundary layers induced by the two horizontal edges. Although this effect rapidly vanishes with increasing scale ratio, neither of the two enriched homogenization schemes is able to capture this phenomenon properly. In the case of the~$10l \times 10l$ specimen, the post-bifurcation stiffness is reproduced accurately by both the second-order as well as the micromorphic method. The first-order method gives, however, the most accurate prediction of the plateau behaviour despite its large error in the predicted bifurcation strain, cf. also Fig.~\ref{fig:ScaleSepFiCom_stiffness_b}.

Graphs of the applied nominal stress are shown in Fig.~\ref{fig:ScaleSepFiCom} as a function of scale ratio for two different nominal strains~$u^{*}/H = 0.01$ (pre-bifurcation) and~$u^{*}/H = 0.05$ (post-bifurcation). From the reported results it may be concluded that the micromorphic homogenization method provides the most accurate or balanced results, although an overestimation of the stress is still observed in Fig.~\ref{fig:ScaleSepFiCom_b}. In the same figure, the second-order method provides overall more accurate result, which is partially due to the choice of the nominal strain at which the nominal stress is evaluated. Pre-bifurcation, a significant overestimation similar to the uniform compression case is observed for second-order scheme.
\begin{figure}[!h]
	\centering
	\subfloat[DNS]{\includegraphics[scale=0.833]{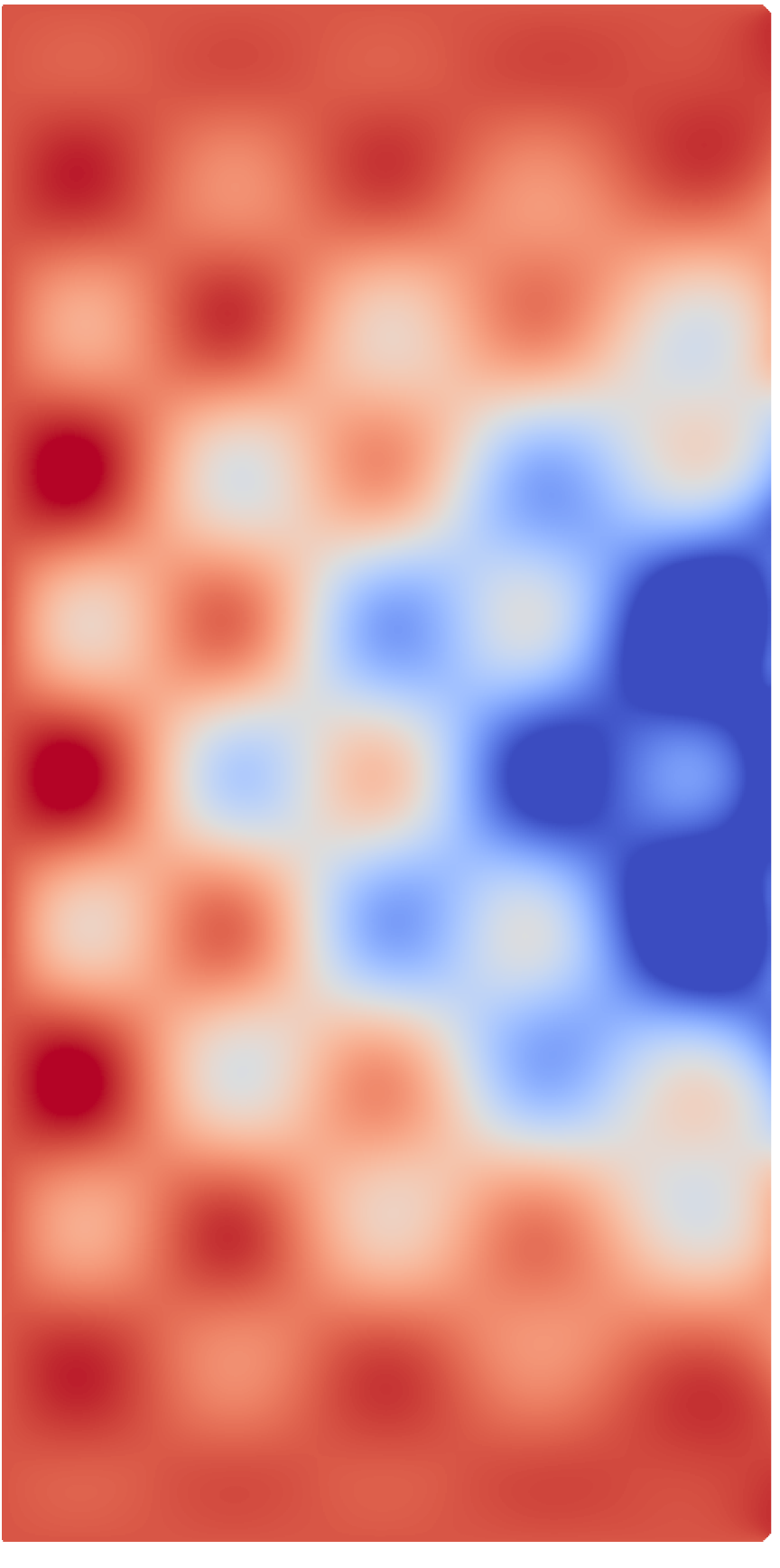}\label{fig:fieldplot_a}}
	\hspace{0mm}
	\subfloat[first-order]{\includegraphics[scale=0.833]{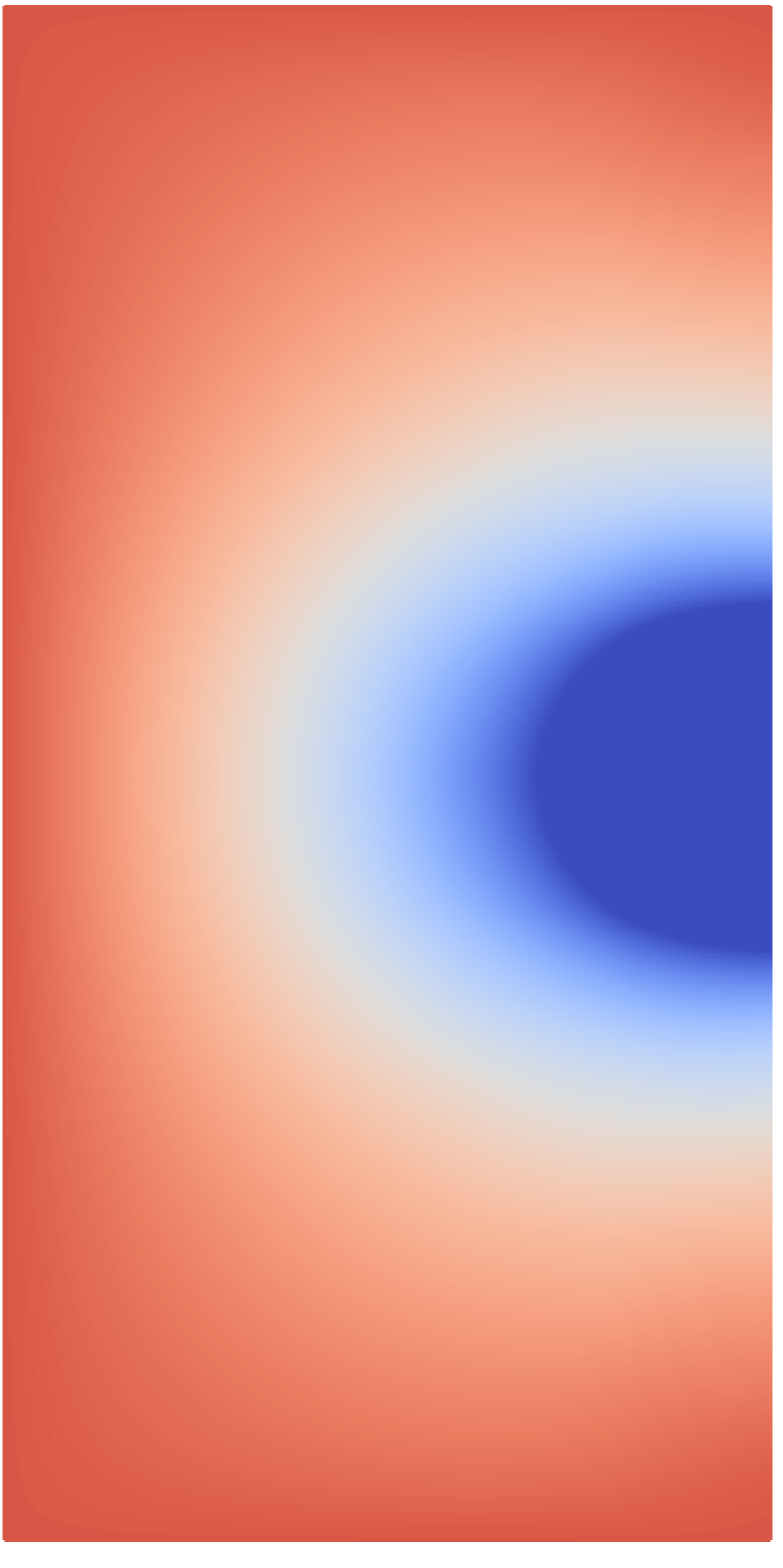}\label{fig:fieldplot_b}}
	\hspace{0mm}
	\subfloat[second-order]{\includegraphics[scale=0.833]{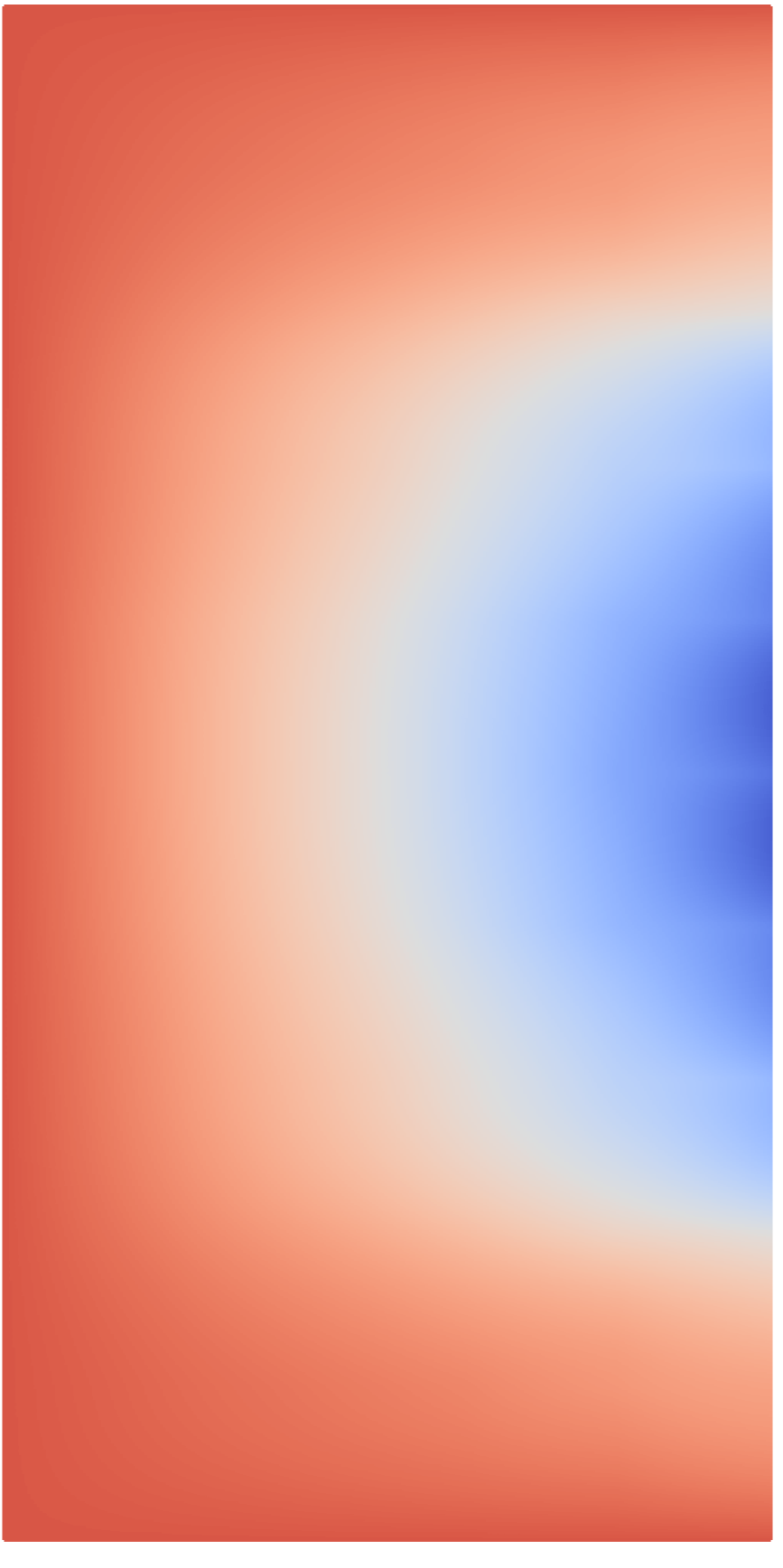}\label{fig:fieldplot_c}}
	\hspace{0mm}
	\subfloat[micromorphic]{\includegraphics[scale=0.833]{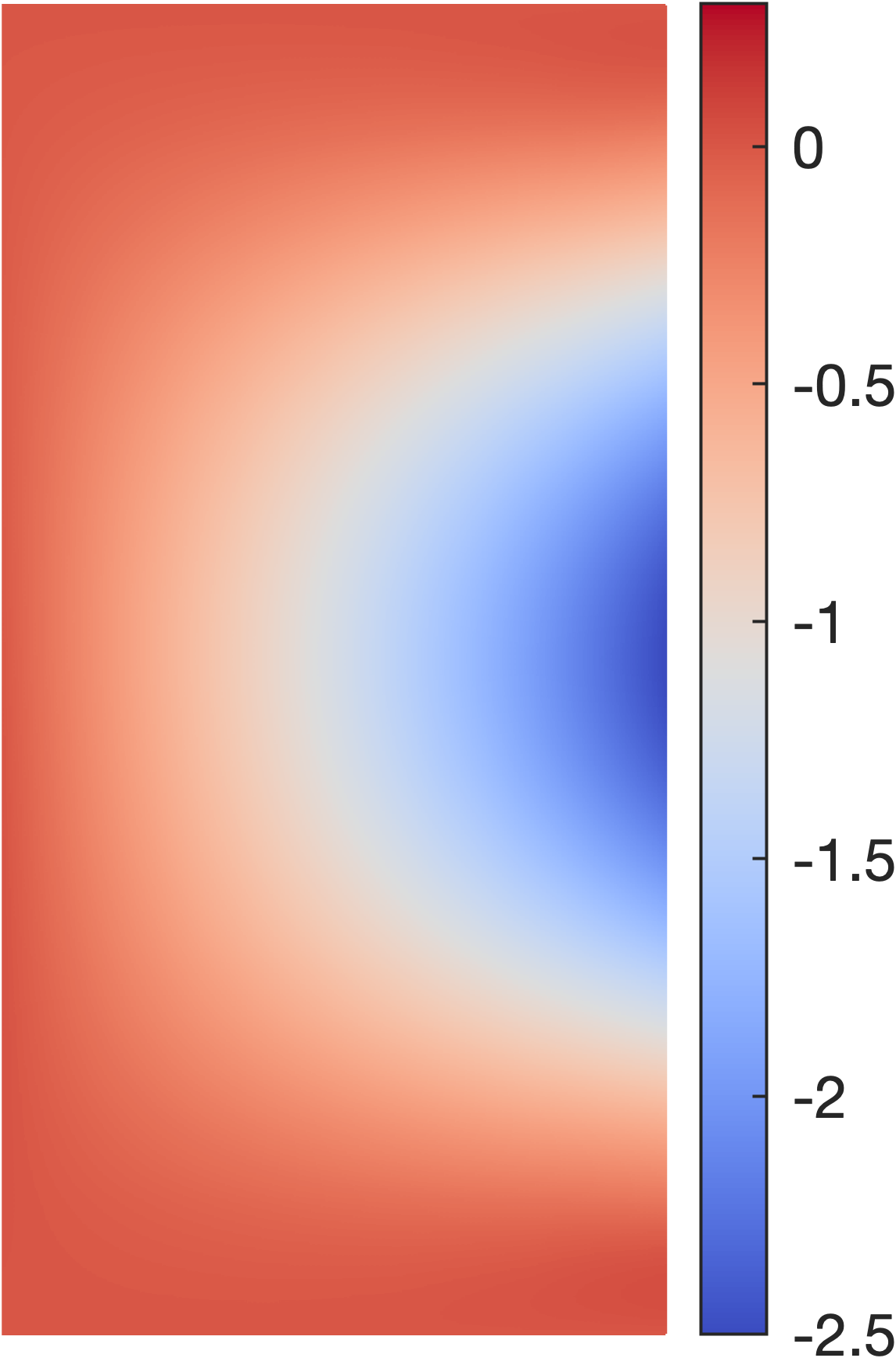}\label{fig:fieldplot_d}}
	\\
	\hspace{0.25mm}
	\subfloat[DNS]{\includegraphics[scale=0.833]{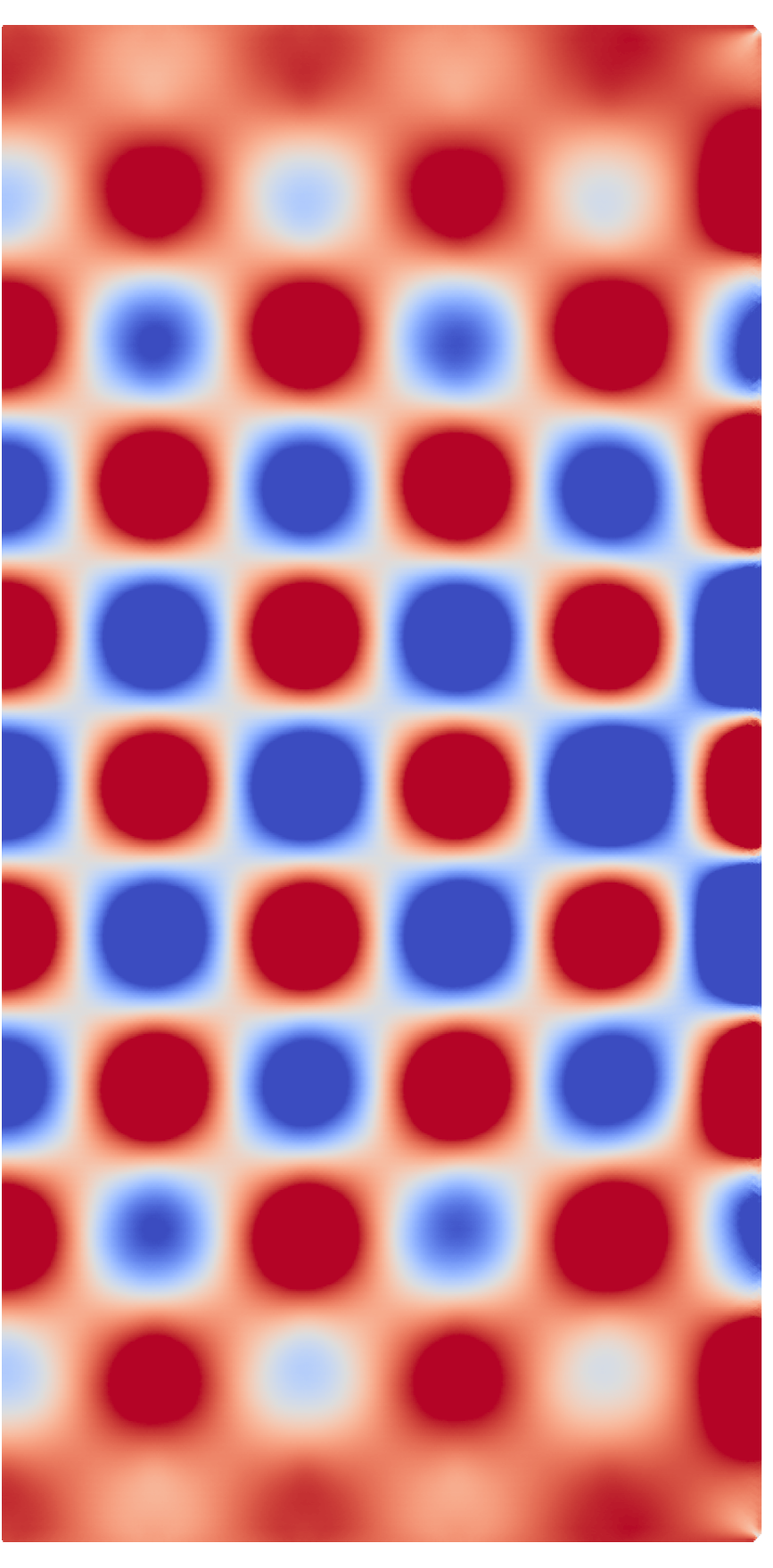}\label{fig:fieldplot_e}}
	\hspace{0mm}
	\subfloat[first-order]{\includegraphics[scale=0.833]{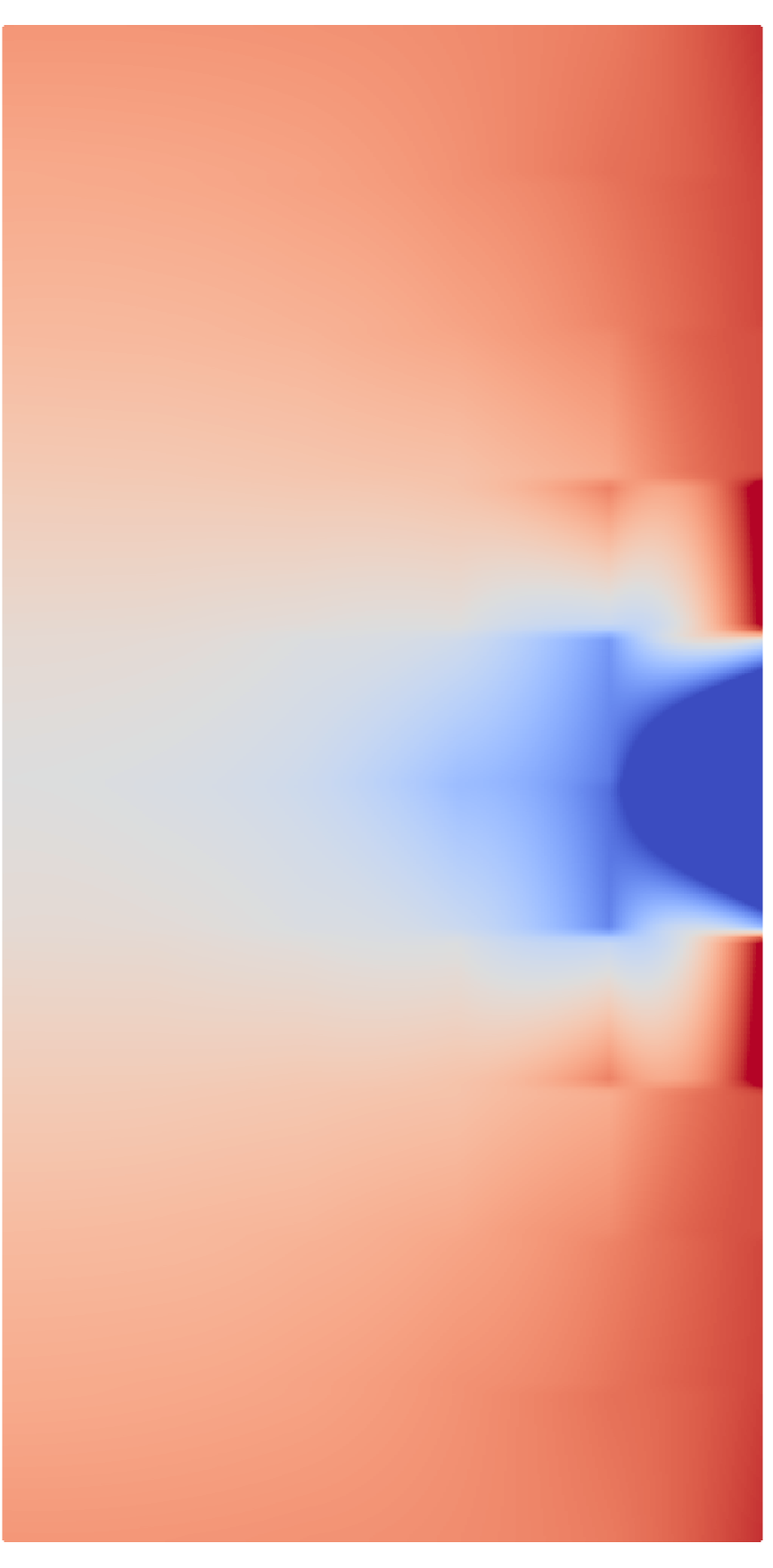}\label{fig:fieldplot_f}}
	\hspace{0mm}
	\subfloat[second-order]{\includegraphics[scale=0.833]{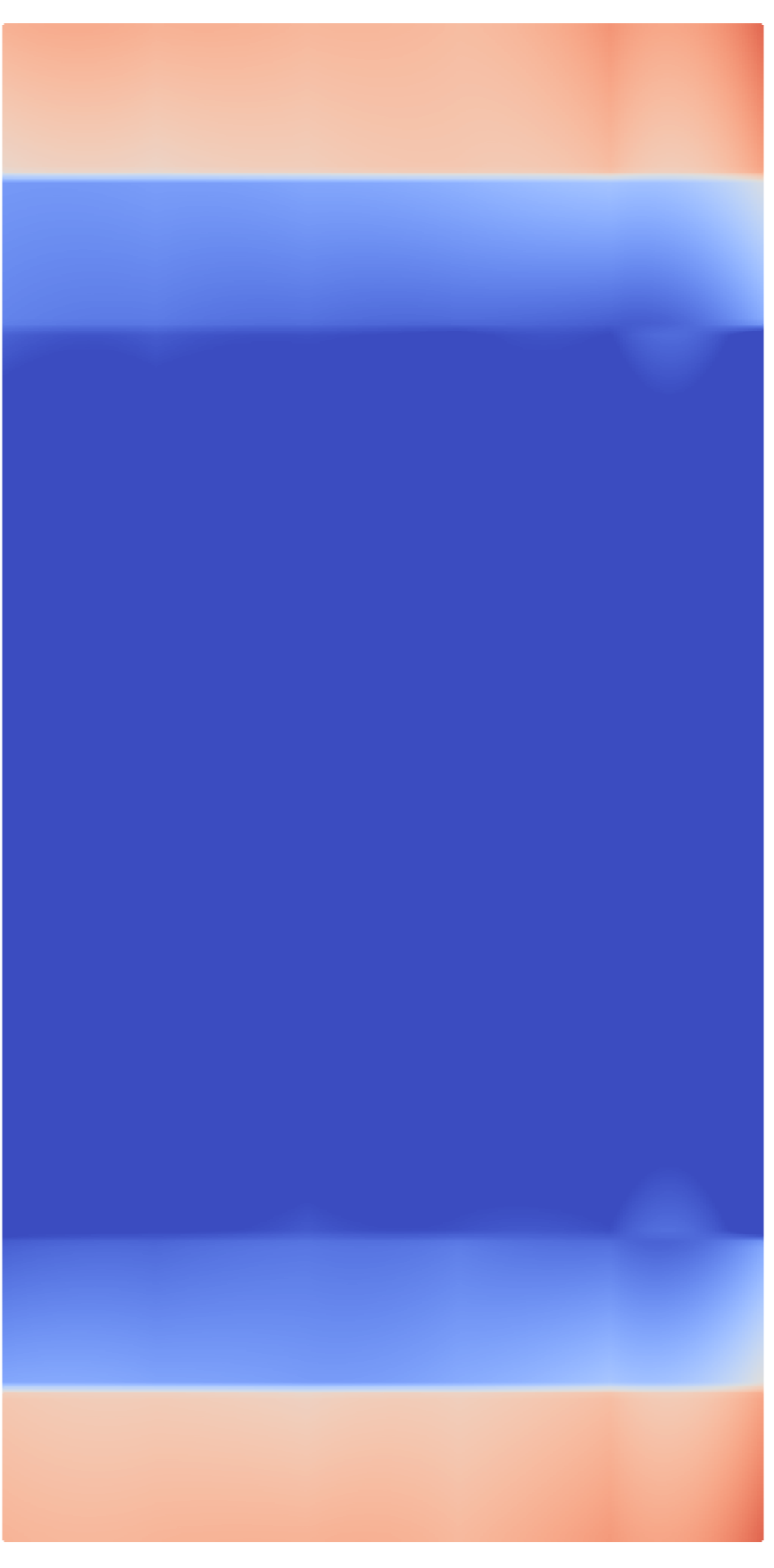}\label{fig:fieldplot_g}}
	\hspace{0mm}
	\subfloat[micromorphic]{\includegraphics[scale=0.833]{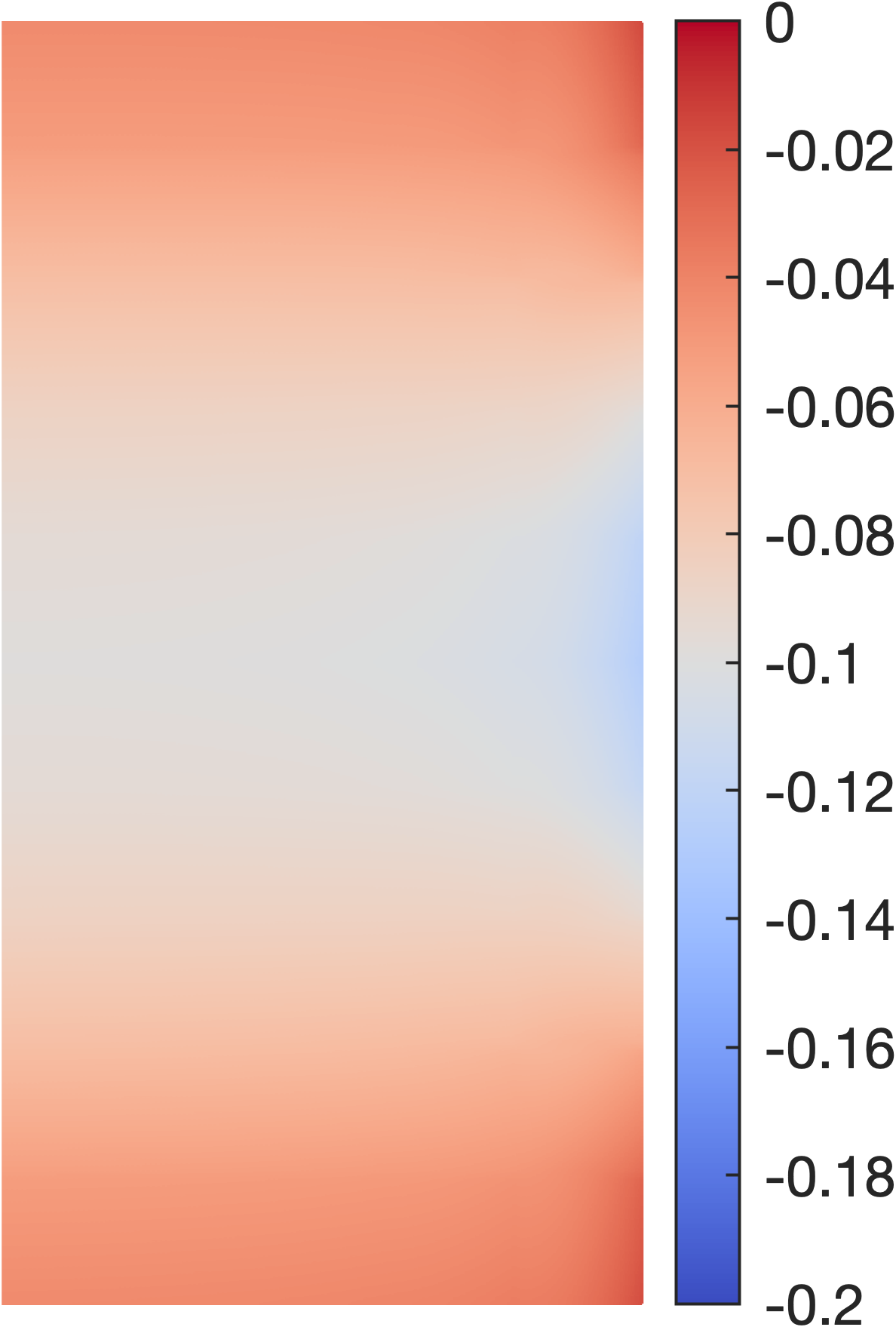}\label{fig:fieldplot_h}}	
	\caption{Finite specimen compression case. Resulting field plots for~$u^{*}/H = 0.075$ and~$H/l = 10$, for the right half of the domain. The top row corresponds to the horizontal displacement component~$\vec{u}\cdot\vec{e}_1$~[mm] for (a)~DNS, (b)~first-order, (c)~second-order, and (d)~micromorphic simulation. The bottom row corresponds to the~$F_{22}-1$~[-] component of the macroscopic deformation for (e)~DNS, (f)~first-order, (g)~second-order, and (g)~micromorphic simulation. All fields are plotted over the reference configuration. DNS ensemble averages are computed for~$21$ shifts~$\vec{\zeta}$ in each direction, showing high fluctuations due to low sampling.}
	\label{fig:fieldplot}
\end{figure}

Fig.~\ref{fig:ScaleSepFiCom_stiffness} shows the tangent stiffness at nominal strains~$u^{*}/H = 0.01$ and~$u^{*}/H = 0.05$ for all adopted scale ratios. In particular, Fig.~\ref{fig:ScaleSepFiCom_stiffness_b} confirms that both enriched homogenization methods fail to capture the softening behaviour observed for the lowest scale ratios considered (cf. Fig.~\ref{fig:FiComStressStrain_a}), and that tangent stiffnesses are significantly overestimated. The micromorphic method is, nevertheless, closest to the reference solution within the evaluated range. Fig.~\ref{fig:ScaleSepFiCom_stiffness_a} is similar to Fig.~\ref{fig:ScaleSepFiCom_a}, hence the same discussion and conclusions apply.

Corresponding field plots of the effective horizontal displacements are shown in Figs.~\ref{fig:fieldplot_a}--\ref{fig:fieldplot_d}, whereas field plots of~$F_{22}-1$ component in Figs.~\ref{fig:fieldplot_e}--\ref{fig:fieldplot_h}. In all cases, the state corresponds to applied nominal strain~$u^{*}/H = 0.075$ in the post-bifurcation regime. The field plots clearly show that the DNS results are under-sampled and exhibit strong fluctuations, in spite of using~$21$ shifts in both horizontal and vertical directions. Nevertheless, a clear auxetic effect can be observed in Fig.~\ref{fig:fieldplot_a}, whereas the average vertical strain component is almost uninformative, although boundary layers of size $3l$ and lower values on average close to the mid-height of the specimen can be observed. For first-order computational homogenization, a clear localization due to separation of scales (i.e., locality) assumption can be observed, in contrast to the two enriched homogenization schemes. Here, the second-order approach shows a slightly more extensive and less localized region of negative displacements as compared to the micromorphic scheme. For the $F_{22}-1$ component, second-order homogenization shows a significantly more localized boundary layer, similarly to the conclusions of Section~\ref{sect:inifinte_compression} in Fig.~\ref{fig:InComDeformationSR6_b}, which is caused by the prescribed value of the deformation gradient $F_{22} = 1$ at the two horizontal edges.
%
%
\subsection{Concluding Remarks}
\begin{table}
	\centering
	\caption{General performance overview of tested computational homogenization methods applied to elastomeric mechanical metamaterials.}
	\label{tab:OverviewMethods}
	\renewcommand{\arraystretch}{1.1}
	\begin{tabular}{l|ccc}
		\multicolumn{1}{c|}{property/method} & first-order & second-order & micromorphic \\ \hline
		pre-bifurcation (no size effect) & \cmark & \xmark & \cmark \\
		post-bifurcation (size effect) & \xmark & \cmark & \cmark \\
		shift of bifurcation point & \xmark & \xmark & \cmark \\
		no prior knowledge & \cmark & \cmark & \xmark
	\end{tabular}
\end{table}
The overall performance of all considered homogenization methods applied to elastomeric mechanical metamaterials, as obtained from the numerical examples, is summarized in Tab.~\ref{tab:OverviewMethods}. In general, it may be concluded that the micromorphic computational homogenization scheme is able to capture most of the present size effects, relying on prior information on the patterning mode~$\vec{\varphi}_1$. The second-order homogenization scheme is less accurate (in particular, it exhibits too strong size effects, especially in the pre-bifurcation regime), but on the other hand is fully general, requiring no prior knowledge.
%
%
\section{Summary and Conclusions}
\label{sec:Conclusion}
In this paper, a thorough comparison of two enhanced computational homogenization schemes applied to pattern-transforming elastomeric mechanical metamaterials has been performed. The first method considered is second-order computational homogenization scheme, which accounts for strain gradients at the macro-scale, and thus includes non-local effects induced by pattern transformations. The second method, a micromorphic computational homogenization scheme, introduces the magnitude of the emerging pattern as an additional macroscopic field, which captures kinematic coupling between individual cells. Whereas the second-order computational homogenization requires solution of a higher-order continuum problem at the macro-scale, the micromorphic scheme adds one coupled scalar partial differential equation next to the classical macroscopic balance equation. Both methodologies have been compared against a reference, ensemble-averaged solution, which is obtained through Direct Numerical Simulations~(DNS) by including a series of translated microstructures. The first-order homogenization method is included as well to reveal the limitations of this current standard in homogenization solutions. Three distinct loading cases were evaluated to demonstrate the accuracy of the enriched methods: uniform compression, bending, and compression of finite specimens. The most important conclusions can be summarized as follows:
\begin{enumerate}
    \item Both enhanced homogenization schemes provide a reasonable estimate of the effective behaviour exhibited by the elastomeric mechanical metamaterials adopted in this study.

    \item The micromorphic homogenization scheme provides more accurate results in terms of the kinematics as well as the stress quantities as compared to second-order computational homogenization scheme.

    \item The deformed Representative Volume Elements~(RVEs) obtained by both methods correspond well with the DNS results due to the presence of the macroscopic strain gradient or a micromorphic field.

    \item Due to very small size effects in the case of bending, both enhanced homogenization methods overestimate the effective stress in the post-bifurcation regime. In the worst case, i.e., for scale ratio~$4$, the second-order and micromorphic method are approximately~$38\%$ and~$15\%$ above the corresponding reference whereas the first-order homogenization method leads to about~$5\%$ of error.
	
    \item The auxetic effect, present in the compression of the finite specimen example, is captured accurately by both enhanced homogenization schemes. The shape of the soft boundary, however, is more accurate for the micromorphic method.

    \item The compression of finite specimens proves to be a challenging task for all computational homogenization schemes, especially in terms of the nominal stress--strain response. Whereas the micromorphic method is still able to capture the bifurcation point with reasonable accuracy, the second-order computational homogenization introduces an error of the order of~$50\%$ for~$4l \times 4l$ specimen. The post-bifurcation stiffness for the same specimen is significantly overestimated, but converges rapidly towards the correct value with increasing scale ratio.
\end{enumerate}

From the presented results, it may be concluded that the micromorphic computational homogenization method proves to be the most suitable numerical tool for homogenization of elastomeric mechanical metamaterials experiencing pattern-transformations, although second-order computational homogenization holds promise as well. Its main drawback is the overly strong non-local effect in the absence of patterning, often leading to overestimation of effective properties.
%
%
\section*{Acknowledgments}
The research leading to these results has received funding from the European Research Council under the European Union's Seventh Framework Programme (FP7/2007-2013)/ERC grant agreement \textnumero~[339392] and from the Czech Science Foundation (GAČR) grant agreement \textnumero~[19-26143X] (O. Rokoš 03/2019–12/2019). The authors would furthermore like to thank Maqsood M. Ameen for numerous fruitful discussions in the early stages of the project, related to micromorphic computational homogenization.
\nocite{NME2579}
%
%
\bibliography{mybibfile}
\end{document}

%% file: lineno_fix.tex
\newcommand*\patchAmsMathEnvironmentForLineno[1]{%
\expandafter\let\csname old#1\expandafter\endcsname\csname #1\endcsname
\expandafter\let\csname oldend#1\expandafter\endcsname\csname end#1\endcsname
\renewenvironment{#1}%
{\linenomath\csname old#1\endcsname}%
{\csname oldend#1\endcsname\endlinenomath}}%
\newcommand*\patchBothAmsMathEnvironmentsForLineno[1]{%
\patchAmsMathEnvironmentForLineno{#1}%
\patchAmsMathEnvironmentForLineno{#1*}}%
\AtBeginDocument{%
\patchBothAmsMathEnvironmentsForLineno{equation}%
\patchBothAmsMathEnvironmentsForLineno{align}%
\patchBothAmsMathEnvironmentsForLineno{flalign}%
\patchBothAmsMathEnvironmentsForLineno{alignat}%
\patchBothAmsMathEnvironmentsForLineno{gather}%
\patchBothAmsMathEnvironmentsForLineno{multline}%
}